\documentclass{aa}  
\usepackage{txfonts}
\usepackage{hyperref}
\usepackage[english]{babel}

\addto\extrasenglish{%
}
\usepackage[dvipsnames]{xcolor}
\colorlet{mylinkcolor}{Maroon}
\colorlet{mycitecolor}{MidnightBlue}
\colorlet{myurlcolor}{MidnightBlue}
\hypersetup{ 
	colorlinks   = true,
	citecolor    = mycitecolor,
	linkcolor    = mylinkcolor,
	urlcolor	= myurlcolor
}

\usepackage{graphicx}
\usepackage{natbib}
\usepackage{pdfpages}
\usepackage{CJK}

\begin{document} 
\begin{CJK}{UTF8}{bkai}

   \title{Extended H$\alpha$ over compact far-infrared continuum in dusty submillimeter galaxies}
   \subtitle{Insights into dust distributions and star-formation rates at $z\sim2$}

   \author{
   	Chian-Chou Chen (陳建州)\inst{1,2}
   	\and
   	C. M. Harrison\inst{3,2}
   	\and
   	I. Smail\inst{4}
   	\and
   	A. M. Swinbank\inst{4}
   	\and
   	O. J. Turner\inst{5}
   	\and 
   	J. L. Wardlow\inst{6}
   	\and
   	W. N. Brandt\inst{7,8,9}
   	\and 
   	G. Calistro Rivera\inst{10,2}
   	\and 
   	S. C. Chapman\inst{11,12,13,14}
   	\and 
   	E. A. Cooke\inst{4}
   	\and 
   	H. Dannerbauer\inst{15,16}
   	\and 
   	J. S. Dunlop\inst{5}
   	\and 
   	D. Farrah\inst{17,18}
   	\and 
   	M. J. Micha\l{}owski\inst{19}
   	\and 
   	E. Schinnerer\inst{20}
   	\and 
   	J. M. Simpson\inst{1}
   	\and 
   	A. P. Thomson\inst{21}
   	\and 
   	P. P. van der Werf\inst{10}
     }
          
   \institute{Academia Sinica Institute of Astronomy and Astrophysics, P.O. Box 23-141, Taipei 10617, Taiwan \\
   	\email{ccchen@asiaa.sinica.edu.tw}
   	\and
	European Southern Observatory, Karl Schwarzschild Strasse 2, Garching, Germany	
   	\and
   	School of Mathematics, Statistics and Physics, Newcastle University, Newcastle upon Tyne, NE1 7RU , UK
   	\and
   	Centre for Extragalactic Astronomy, Department of Physics, Durham University, South Road, Durham DH1 3LE, UK
   	\and
   	Institute for Astronomy, University of Edinburgh, Royal Observatory, Blackford Hill, Edinburgh EH9 3HJ, UK
   	\and
   	Physics Department, Lancaster University, Lancaster, LA14YB, UK
   	\and
   	Department of Astronomy \& Astrophysics, 525 Davey Lab, The Pennsylvania State University, University Park, PA 16802, USA
   	\and
   	Institute for Gravitation and the Cosmos, The Pennsylvania State University, University Park, PA 16802, USA
   	\and
   	Department of Physics, The Pennsylvania State University, University Park, PA 16802, USA
   	\and
   	Leiden Observatory, Leiden University, P.O. Box 9513, 2300 RA Leiden, the Netherlands
   	\and
   	Department of Physics \& Astronomy, University of Victoria, BC, V8X 4M6, Canada
   	\and
   	Department of Physics and Atmospheric Science, Dalhousie University, Halifax, NS, B3H 4R2, Canada
   	\and
   	NRC Herzberg Astronomy and Astrophysics, 5071 West Saanich Road, Victoria, BC, V9E 2E7, Canada
   	\and
   	Department of Physics and Astronomy, University of British Columbia, Vancouver, BC, V6T 1Z1, Canada
   	\and
   	Instituto de Astrofísica de Canarias (IAC), E-38205 La Laguna, Tenerife, Spain
   	\and
   	Universidad de La Laguna, Dpto. Astrofísica, E-38206 La Laguna, Tenerife, Spain
   	\and
   	Department of Physics and Astronomy, University of Hawaii, 2505 Correa Road, Honolulu, HI 96822, USA
   	\and
   	Institute for Astronomy, 2680 Woodlawn Drive, University of Hawaii, Honolulu, HI 96822, USA
   	\and
   	Astronomical Observatory Institute, Faculty of Physics, Adam Mickiewicz University, ul.~S{\l}oneczna 36, 60-286 Pozna{\'n}, Poland
   	\and
   	Max–Planck Institut f\"{u}r Astronomie, K\"{o}nigstuhl 17, 69117 Heidelberg, Germany
   	\and
   	The University of Manchester, Oxford Road, Manchester, M139PL, UK 
   }
   
   \date{Accepted by A\&A}

  \abstract{Using data from ALMA and near-infrared (NIR) integral field spectrographs including both SINFONI and KMOS on the VLT, we investigate the two-dimensional distributions of H$\alpha$ and rest-frame far-infrared (FIR) continuum in six submillimeter galaxies at $z\sim2$. At a similar spatial resolution ($\sim$0$\farcs$5 FWHM; $\sim$4.5\,kpc at $z=2$), we find that the half-light radius of H$\alpha$ is significantly larger than that of the FIR continuum in half of the sample, and on average H$\alpha$ is a median factor of $2.0\pm0.4$ larger. Having explored various ways to correct for the attenuation, we find that the attenuation-corrected H$\alpha$-based SFRs are systematically lower than the IR-based SFRs by at least a median factor of $3\pm1$, which cannot be explained by the difference in half-light radius alone. In addition, we find that in 40\% of cases the total $V$-band attenuation ($A_V$) derived from energy balance modeling of the full ultraviolet(UV)-to-FIR spectral energy distributions (SEDs) is significantly higher than that derived from SED modeling using only the UV-to-NIR part of the SEDs, and the discrepancy appears to increase with increasing total infrared luminosity. Finally, considering all our findings along with the studies in the literature, we postulate that the dust distributions in SMGs, and possibly also in less IR luminous $z\sim2$ massive star-forming galaxies, can be decomposed into three main components; the diffuse dust heated by older stellar populations, the more obscured and extended young star-forming H{\sc ii} regions, and the heavily obscured central regions that have a low filling factor but dominate the infrared luminosity in which the majority of attenuation cannot be probed via UV-to-NIR emissions.}

   \authorrunning{C.-C. Chen et al.}
   \maketitle

%

\section{Introduction}
	
	Measurements of star-formation rate (SFR) across cosmic time provide one of the most fundamental constraints to models of galaxy formation and evolution (e.g., \citealt{Somerville:2015aa}). Comparisons between SFR and other galaxy properties have yielded essential insights into the physics of galaxy assembly, such as the Schmidt-Kennicutt relationship \citep{Schmidt:1959aa,Kennicutt:1989aa} and the so-called galaxy star-forming main sequence (e.g., \citealt{Noeske:2007fk,Elbaz:2011aa,Whitaker:2012aa,Schreiber:2015aa}). 
	
	The vital role that SFR measurements play in constraining galaxy models means that the calibrations and diagnostics for various tracers have been extensively studied \citep{Kennicutt:2012aa}. Thanks to technical advances in both ground-based telescopes and space-based satellites, SFRs can be estimated through a wide spectral range of emissions from X-ray to radio, using both continuum and line emission (e.g., \citealt{Ranalli:2003aa,Lehmer:2010aa,Murphy:2011aa}). Since measurements made at different wavelengths are sensitive to different ages of the stellar populations \citep{Calzetti:2013aa}, and they are affected by dust attenuation to a different level, in principle SFR tracers at different wavelengths should all be considered and exploited for their complementary strengths of diagnostics. Indeed in the local universe, calibrations have been derived to obtain the total SFRs, in particular in addressing dust attenuation, by combing UV, H$\alpha$, and infrared data \citep{Hao:2011aa}. This has been done both locally for individual star-forming regions \citep{Calzetti:2007aa,Li:2013aa} and globally for entire galaxies \citep{Kennicutt:2009aa,Catalan-Torrecilla:2015aa,Brown:2017aa}.

	Beyond the nearby universe, the volume-averaged SFR density rises rapidly, by over an order of magnitude by $z\sim2$ \citep{Madau:2014aa}. During these times the SFR density is dominated by infrared bright galaxies \citep{Le-Floch:2005fk,Smolcic:2009aa,Gruppioni:2013aa,Swinbank:2014aa}, which are classified as (Ultra-)luminous infrared galaxies ((U)LIRGs; \citealt{Sanders:1996p6419}) with total infrared luminosity greater than ($10^{12}$)$10^{11}$\,L$_\odot$. This means that at $z\sim2$ the infrared component of the SFRs becomes dominant, and the correction of dust attenuation becomes critical for galaxy samples in which the SFRs are predominantly measured via UV and optical. Indeed, continuum and emission lines in the UV and optical are extensively used at high redshifts to measure SFRs given their accessibility and technical advances, and the cosmic SFR density has been estimated with this method up to $z\sim10$ (e.g, \citealt{Reddy:2009aa,Bouwens:2011vn,Ellis:2013vn,Oesch:2015aa,McLeod:2016aa,Ishigaki:2018aa}).

	  \begin{table*}
	  	\caption{General information of the sample and the data}             
	  	\label{table1}      
	  	\begin{center}
	  		\begin{tabular}{lrrlrrllr}
	  			\hline
	  			ID & R.A.$_{\text{ALMA}}$ & Decl.$_{\text{ALMA}}$ & AGN$^a$ & PSF$_{\text{ALMA,nat}}$ & PSF$_{\text{ALMA,tap}}$ & IFU & IFU$_{\text{band}}$ & PSF$_{\text{IFU}}$\\
	  			& [J2000;degree] & [J2000;degree] &  & [arcsecond] & [arcsecond] &  &  & [arcsecond]\\
	  			\hline
	  			ALESS17.1 & 53.030410 & $-$27.855765 & X-ray & 0.17$\times$0.15 & 0.66$\times$0.62 & SINFONI & H & 0.65\\
	  			ALESS66.1 & 53.383053 & $-$27.902645 & No & 0.17$\times$0.13 & 0.47$\times$0.45 & SINFONI & K & 0.48\\
	  			ALESS67.1 & 53.179981 & $-$27.920649 & X-ray & 0.18$\times$0.15 & 0.66$\times$0.62 & SINFONI & HK & 0.66\\
	  			ALESS75.1 & 52.863303 & $-$27.930928 & IR & 0.17$\times$0.12 & 0.57$\times$0.54 & SINFONI & HK & 0.56\\
	  			AS2UDS292.0 & 34.322638 & $-$5.2300513 & X-ray & 0.22$\times$0.19 & 0.46$\times$0.42 & KMOS & K & 0.46\\
	  			AS2UDS412.0 & 34.422392 & $-$5.1810288 & No & 0.25$\times$0.23 & 0.56$\times$0.52 & KMOS & K & 0.55\\
	  			\hline
	  		\end{tabular}
	  		\tablefoottext{a}{The AGN classification in X-ray is based on catalog matching; ALESS SMGs are matched to the {\it Chandra} 7\,Ms catalog and classified as AGN based on the criteria set in \citet{Luo:2017aa}. The AS2UDS SMGs are matched to the {\it Chandra} X-UDS catalog \citep{Kocevski:2018aa} with a simple luminosity cut of $L_{\rm 2-10kev} > 3\times10^{42}$\,erg s$^{-1}$, which is one of the criteria used in \citet{Luo:2017aa}. The steep slope in mid-infrared has suggested that ALESS75.1 is an IR AGN \citep{Simpson:2014aa,Stanley:2018aa}.}
	  	\end{center}
	  \end{table*}
	  
	However the method to correct for dust attenuation for UV and optical measurements at high redshifts is currently a topic of debate. For example, the common method to correct the UV SFRs using the correlation between the ratio of infrared and UV luminosity and the spectral slope in the UV, the IRX-$\beta$ relation, is subject to many systematics including turbulence, the age of the stellar populations, and the dust compositions and geometry (e.g., \citealt{Howell:2010aa,Casey:2014ab,Chen:2017aa,Popping:2017aa,Narayanan:2018aa}). Similarly, due to the faintness of H$\beta$, the correction for the H$\alpha$-based SFRs using the Balmer decrement is also not straightforward. To bypass the difficulty of obtaining the Balmer decrement, stellar attenuation derived from SED fitting based on UV-to-NIR photometry has been adopted (e.g., \citealt{Sobral:2013aa}). However the relation between the nebular attenuation and the stellar attenuation has not been well determined, both locally (e.g., \citealt{Kreckel:2013aa}) and at high redshifts (e.g., \citealt{Calzetti:2000aa,Wild:2011aa,Kashino:2013aa,Price:2014aa,Reddy:2015aa,Puglisi:2016aa,Theios:2018aa}).
	
	Regardless of whether the galaxies are located in the nearby universe or at high redshifts, one of the main issues faced when correcting dust attenuation has been the spatial distribution of dust. It is assumed that dust acts as a foreground screen and spatially coincides with the underlying emissions, which is partly motivated by the findings in the local spiral galaxies (e.g., \citealt{Kennicutt:2009aa,Bendo:2012aa}). In addition, motivated by often times different attenuations found in nebular emission lines such as H$\alpha$ compared to the stellar continuum (e.g., \citealt{Calzetti:2000aa}), the dust distribution is normally perceived as having two main components; the diffuse dust in the interstellar medium (ISM) that obscured older stellar populations and the dustier component obscuring the H{\sc ii} regions (e.g., \citealt{Wild:2011aa, Price:2014aa, Reddy:2015aa, Leslie:2018aa}). This two-component dust distribution model is one of the major assumptions that goes into the SED modeling that employs the energy balance technique, such as {\sc magphys} \citep{da-Cunha:2008aa,da-Cunha:2015aa} and {\sc cigale} \citep{Noll:2009aa}.
 
    \begin{figure*}[thb]
    	\centering
    	\includegraphics[scale=0.6]{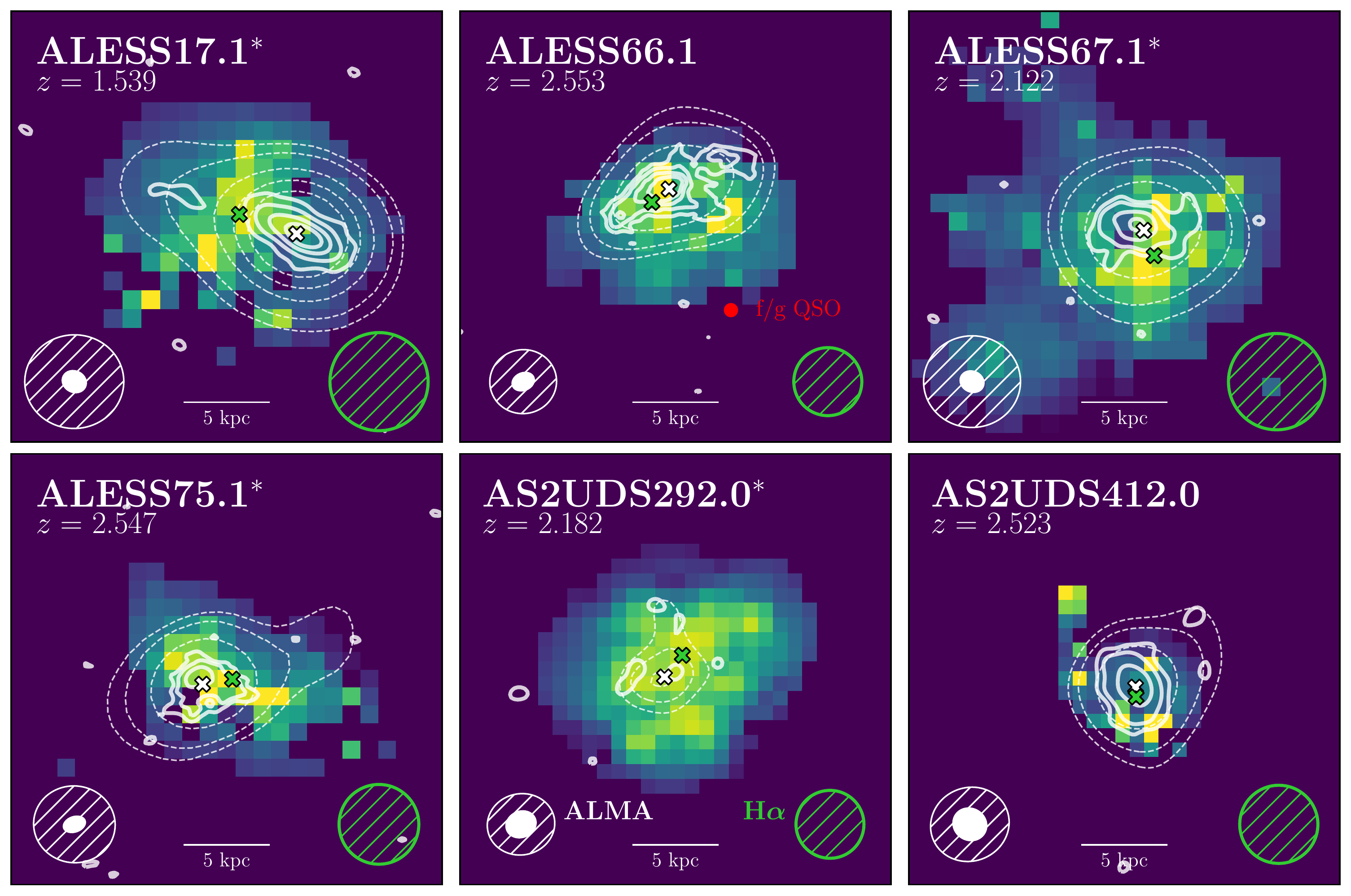}
    	\caption{Two dimensional H$\alpha$ maps of our sample SMGs (\autoref{table1}) with a size of 25$\times$25\,kpc and a color stretch from zero to 99.5\% of the peak. The solid and the dashed contours starting from 3\,$\sigma$ are overlaid on top to show two versions of the ALMA 870\,$\mu$m continuum, respectively; One produced using natural weighting, resulting in $\sim$0\farcs2 resolution, and the other tapered to $\sim$0\farcs5 resolution, matching to the resolution of the H$\alpha$ images. The resolution beams in sizes of FWHM are plotted in the bottom corners and the exact sizes are given in \autoref{table1}. {White and green crosses mark the positions adopted as the centers of the curve-of-growth analyses, for 870\,$\mu$m continuum and H$\alpha$, respectively. These positions are centroids of the best-fit ellipses to the isophotes of the corresponding 2D images, with the details described in \autoref{sec:analyses}.} Sources that are identified to host an AGN are marked with an asterisk after the ID. The nearby foreground quasar next to ALESS66.1 is marked as a red dot. We find that the rest-frame FIR emissions appear to be smaller in scale than H$\alpha$ in most sources.}
    	\label{fig1}%
    \end{figure*}
	  
	In more chaotic and gas-rich environments, in particular at $z\sim2$, there is increasing evidence suggesting that the aforementioned assumptions may need to be adjusted. For example, studies using ALMA have found, almost ubiquitously, that the massive dusty galaxies at high redshifts have very compact morphology in FIR continuum, with half-light radius of $\sim1-2$\,kpc (e.g., \citealt{Simpson:2015aa,Ikarashi:2015aa,Ikarashi:2017aa,Barro:2016aa,Harrison:2016ab,Hodge:2016aa,Hodge:2019aa,Spilker:2016aa,Tadaki:2017aa,Oteo:2017aa,Fujimoto:2018aa}). In addition, many other tracers of star-forming regions are found to be spatially offset from, or much larger than, the FIR continuum, including the stellar continuum in UV/optical (e.g., \citealt{Chen:2015aa,Hodge:2016aa,Cowie:2018aa,Gomez-Guijarro:2018aa}), radio continuum at 1.4 and 3\,GHz (e.g., \citealt{Biggs:2008fj,Miettinen:2017aa,Thomson:2019aa}), and emission lines such as [CII] \citep{Gullberg:2018aa,Litke:2018aa}, $^{12}$CO \citep{Spilker:2015aa,Chen:2017aa,Tadaki:2017ab,Calistro-Rivera:2018aa,Dong:2019aa}, H$_2$O \citep{Apostolovski:2019aa}, and H$\alpha$ \citep{Alaghband-Zadeh:2012aa,Menendez-Delmestre:2013aa,Chen:2017aa,Nelson:2019aa}. While these studies mostly focused on more IR luminous sources, these results suggest that the dust distribution at $z\sim2$ could be significantly different than that in galaxies in nearby universe, and spatially resolved studies comparing various star formation tracers are needed in order to better understand the physics of star formation during the epoch when most of the massive elliptical galaxies seen in the nearby universe are formed (e.g., \citealt{Lilly:1999lr, Hickox:2012kk, Toft:2014aa, Simpson:2014aa, Chen:2016ac, Wilkinson:2017aa}). 
	
	This paper is motivated by \citet{Chen:2017aa}, in which we found that in a submillimeter galaxy (SMG), ALESS67.1, at $z=2.12$ where we managed to gather sub-arcsecond UV-to-NIR continuum, FIR continuum, $^{12}$CO, and H$\alpha$, the size of FIR continuum is a factor of $\sim$4-6 smaller than that of all the other emissions. \citet{Calistro-Rivera:2018aa} have extended the size comparison between FIR continuum and $^{12}$CO to a sample of four SMGs, finding that $^{12}$CO($J=3-2$) is larger than the FIR continuum by a factor of $>2$. They propose that the size difference can be explained by temperature and optical depth gradients alone. In this paper we aim to extend the comparison between the FIR continuum and H$\alpha$ to a larger sample of SMGs. In \autoref{sec:obs} we provide details of our sample selection and data. The analyses and measurements are presented in \autoref{sec:analyses}. We discuss the implications of our findings in \autoref{sec:results} and the summary is given in \autoref{sec:sum}. Throughout this paper we define the size to be the half-light radius. We assume the {\it Planck} cosmology: H$_0 =$\,67.8\,km\,s$^{-1}$ Mpc$^{-1}$, $\Omega_M = $\,0.31, and $\Omega_\Lambda =$\,0.69 \citep{Planck-Collaboration:2014aa}. 

    \begin{figure*}[thb]
    	\centering
    	\includegraphics[scale=0.74]{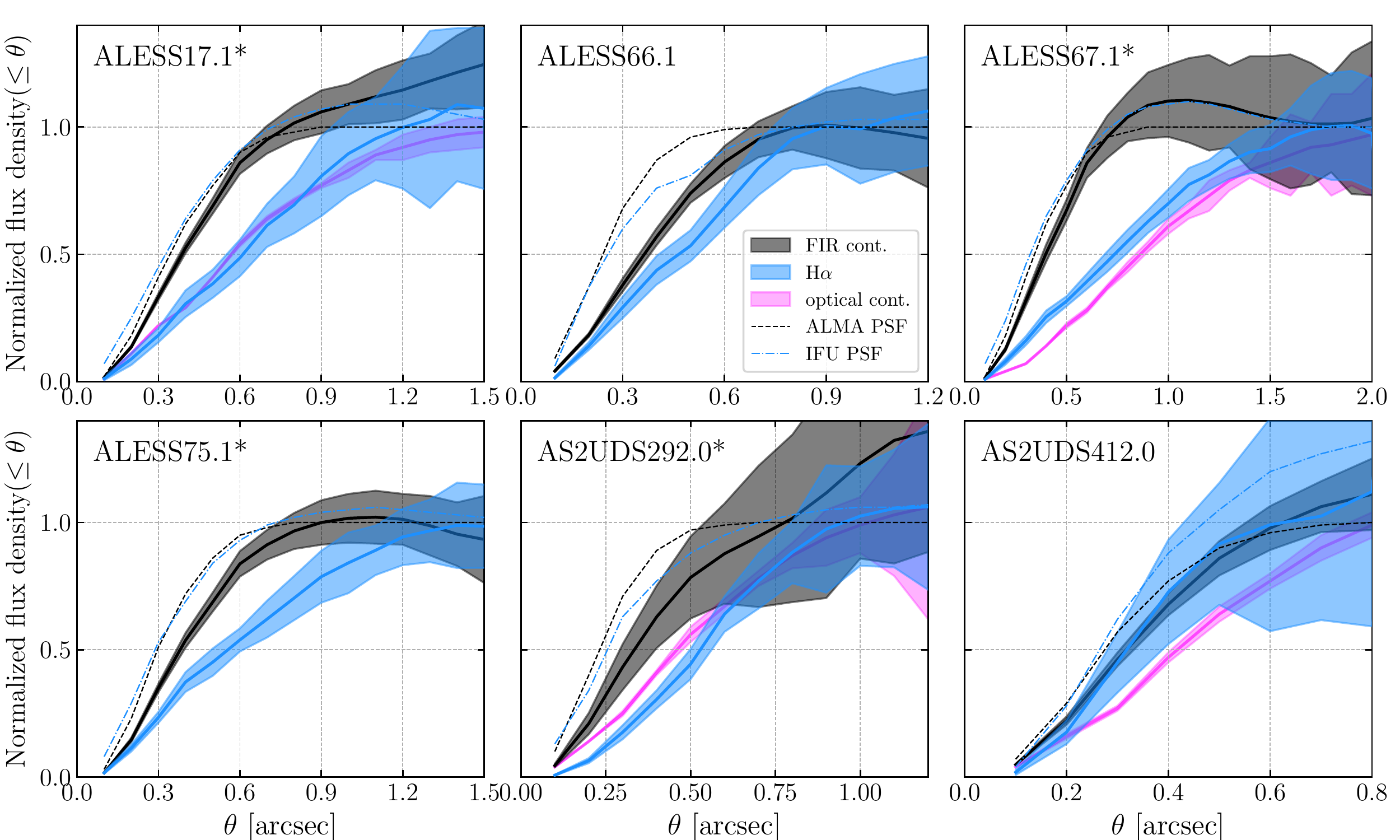}
    	\caption{A plot of curve-of-growth for each of our sample SMGs, showing the rest-frame FIR continuum in black, H$\alpha$ in blue, and wherever available rest-frame optical continuum from {\it HST} in magenta. {Sources that are identified to host an AGN are marked with an asterisk after the ID.} The flux densities are normalized to the total flux densities, and the 1\,$\sigma$ uncertainties are shown as respective color bands. All FIR and optical continuum images are convolved to a resolution matched to that of the H$\alpha$ images. The curve-of-growth results for the PSFs are also shown as dashed curves, {which are based on the synthesized beam for ALMA and the standard stars for the IFU data}.}
    	\label{fig2}%
    \end{figure*}
    
\section{Sample, data, and reductions}\label{sec:obs}
\subsection{Sample}
Our sample of six SMGs is drawn from two parent SMG samples. One is the ALESS sample \citep{Hodge:2013lr,Karim:2013fk}, obtained from a Cycle 0 ALMA 850\,$\mu$m follow-up survey targeting a flux-limited sample of 126 submillimeter sources detected by a LABOCA \citep{Siringo:2009rt} 870\,$\mu$m survey in the Extended {\it Chandra} Deep Field South (ECDFS) field (LESS survey; \citealt{Weis:2009qy}). These sources were subsequently observed at a higher angular resolution with ALMA \citep{Hodge:2016aa,Hodge:2019aa}. The other is the AS2UDS sample \citep{Simpson:2017ab,Stach:2018aa,Stach:2019aa}, based on a Cycle 1,3,4,5 ALMA 850\,$\mu$m follow-up program of all 716 $>4$\,$\sigma$ submillimeter sources uncovered by the SCUBA-2 850\,$\mu$m legacy survey in the UKIDSS-UDS field \citep{Geach:2017aa}. 

To compare the morphology of FIR continuum and H$\alpha$, in particular the measurement of sizes, it requires spatially resolved observations, which for SMGs typically means observations taken at $\lesssim$0$\farcs$5 (e.g., \citealt{Simpson:2015aa,Alaghband-Zadeh:2012aa}). Therefore for the selection of galaxies from ALESS and AS2UDS for this study we require, firstly, that targets have $\lesssim$0$\farcs$5 angular resolution ALMA band 7 continuum imaging, which has a matched or better spatial resolution to the seeing-limited H$\alpha$ data so it allows direct comparisons in spatial distributions between cold dust continuum and H$\alpha$. Majority of the AS2UDS SMGs satisfy this criterion \citep{Stach:2019aa} and for ALESS SMGs we consider the ones published in \citet{Hodge:2016aa} and an extra sample obtained through the program 2016.1.00735.S (PI: C. M. Harrison).

The second step is to based on the positions of these SMGs search the VLT archive for the existing IFU (SINFONI or KMOS) observations. We reduce the archival data following the methods described in the next section. To allow creations of two-dimensional intensity maps we keep the SMGs that have high signal-to-noise ratio H$\alpha$ detection (but remove obvious broad line active galactic nuclei; BLAGN), corresponding to a typical flux density of $\sim10^{-16}$\,erg s$^{-1}$ cm$^{-2}$. While the archival data have been taken by different programs so it is difficult to assess the potential selection biases, we note that the H$\alpha$ flux density distribution ($5\times10^{-17} - 3\times10^{-16}$\,erg s$^{-1}$ cm$^{-2}$) of our sample SMGs follows that reported in the literature on other SMG samples \citep{Swinbank:2004aa,Casey:2017aa}. Under these two criteria we have obtained four ALESS and two AS2UDS SMGs, and their basic properties are given in \autoref{table1}. 

{Based on the SED analyses shown later using the {\sc magphys} code, we find that these six SMGs have a median dust mass of log(M$_{\rm dust}$)=8.9\,$M_\odot$, a median stellar mass of log(M$_\ast$)=11.2\,$M_\odot$ and a median SFR of log(SFR)=2.4\,M$_\odot$\,yr$^{-1}$, meaning they are on average located on the upper part of the massive end of the SFR-M$_\ast$ main sequence at $z\sim2$, which is consistent with the behaviour of the general SMG population \citep{da-Cunha:2015aa}.}



\subsection{SINFONI and KMOS IFU}
The SINFONI-IFU data were taken for the four ALESS sources between October 2013 and December 2014 under the program IDs 091.B-0920 and 094.B-0798. We used $H$-, $K$-, and $HK$-band to observe ALESS17.1, ALESS66.1, and both ALESS67.1 and ALESS75.1, respectively, covering the [N\,{\sc ii}]/H$\alpha$ lines in all four sources and the [O\,{\sc iii}]/H$\beta$ lines in the last two. The spectral resolution is $R > 1500$, sufficient to separate H$\alpha$ and the two [N\,{\sc ii}] lines. The data were reduced using the {\sc esorex} (ESO Recipe Execution Tool) \citep{Freudling:2013aa} pipeline, with additional custom routines applied to improve the sky subtraction. Solutions for flux calibration were derived using the {\sc iraf} routines {\sc standard}, {\sc sensfunc}, and {\sc calibrate} on the standard stars, which normally were observed within two hours of the science observations and processed along with the science data. Standard stars are also used to produce the point spread functions (PSFs) \footnote{The ideal way to monitor the PSF for SINFONI is to request the PSF standard observations but they do not exist in any of the data set used. However we checked the recorded seeing conditions between the standard stars and the science targets and found they agree to $\sim$15\% without a significant systematic offset.}, which are fit with 2D Gaussian models to derive the angular resolution. 

The $K$-band KMOS data on AS2UDS SMGs were taken by the KMOS$^{\rm 3D}$ survey \citep{Wisnioski:2015aa} under the program IDs 093.A-0079 and 096.A-0025. The data reduction primarily made use of SPARK (Software Package for Astronomical Reduction with KMOS; \citealt{Davies:2013aa}), implemented using the {\sc esorex}. In addition to the SPARK recipes, custom {\sc python} scripts were run at different stages of the pipeline and are described in detail in \citet{Turner:2017aa}. Standard star observations were carried out on the same night as the science observations and were processed in an identical manner to the science data, which were used for flux calibration and PSF generation. Sky subtraction was enhanced using the SKYTWEAK option within SPARK \citep{Davies:2007aa}, which counters the varying amplitude of OH lines between exposures by scaling families of OH lines independently to match the data. 

The astrometry of both the SINFONI and the KMOS data was corrected by aligning the continuum of the IFU cube to the corresponding ground-based imaging, including TENIS $K$-band \citep{Hsieh:2012aa} or MUSYC $H$-band in CDF-S \citep{Taylor:2009aa}, and UKIDSS $K$-band \citep{Lawrence:2007aa} in UDS, which was first aligned to the GAIA DR2 catalog \citep{Gaia-Collaboration:2018aa} based on sources with $R = 15-20$\,mag (in AB). By doing so we found a significant systematic shift to the north in declination of the ground-based imaging in CDF-S with respect to the GAIA sources by $\sim$0$\farcs$3$\pm$0$\farcs$1, consistent with similar findings in the literature (e.g., \citealt{Xue:2011aa,Dunlop:2017aa,Luo:2017aa,Scholtz:2018aa}). No systematic offset is found in UDS imaging. Assuming the ALMA astrometry aligns with GAIA DR2, the precision of astrometry of the IFU data derived through this exercise is $<0\farcs$2 (around one pixel of the IFU data).   

\subsection{ALMA 870\,$\mu$m continuum}
The ALMA data of ALESS17.1, ALESS67.1 and the two AS2UDS SMGs have been published and described in detail in \citet{Hodge:2016aa}, \citet{Chen:2017aa}, and \citet{Stach:2018aa}. To present the data self-consistently, we re-analyse the data by first creating the calibrated measurement set using the pipeline reduction scripts provided by the ALMA archive, with a corresponding {\sc casa} version used to generate the scripts. ALESS66.1 and ALESS75.1 were observed in Cycle 4 as a comparison sample for testing the size difference between SMGs with or without detectable AGN (Project ID: 2016.1.00735.S). We again use the pipeline reduction script to create the calibrated measurement set under {\sc casa} version 4.7.2. All ALMA data were tuned to the default band 7 continuum observations centered at 344\,GHz/870\,$\mu$m, with 4 $\times$128 dual polarization channels over the 8 GHz bandwidth. At this frequency, ALMA has a 17$\farcs$3 primary beam in FWHM. 

We then make two sets of images; One is created using natural weighting and the other tapered to a spatial resolution matched to the H$\alpha$ data. Both sets of images are deconvolved using the {\sc clean} algorithm, and circular regions with 1$\farcs$5 radius centered at the SMGs are cleaned down to 2\,$\sigma$. The typical angular resolution under natural weighting is $\sim$0$\farcs$2, and $\sim0\farcs5$ for the tapered images (\autoref{table1}), and the corresponding depths in r.m.s. are 30-70\,$\mu$Jy beam$^{-1}$ and 60-300\,$\mu$Jy beam$^{-1}$, respectively.

\begin{table*}
	\caption{FIR continuum and H$\alpha$ measurements of our sample}             
	\label{table2}      
\begin{center}
	\begin{tabular}{lrrrrrrrr}
		\hline
		ID & redshift & $F_{\text{870}}^a$ & $F_{\text{Ha}}^b$ & $r_{\text{e,maj,870,uv}}^c$ & $r_{\text{e,870,cog}}^d$ & $r_{\text{e,maj,Ha,galfit}}^e$ & $r_{\text{e,Ha,cog}}^f$ & $r_{\text{e}}$ ratio$^g$\\
		&  & [mJy] & [1E-16 erg s$^{\text{-1}}$ cm$^{\text{-2}}$] & [kpc] & [kpc] & [kpc] & [kpc] & \\
		\hline
			ALESS17.1 & 1.5392(2) & 8.2(0.2) & 0.6(0.2) & 1.7(0.1) & 1.7(0.2) & 4.3(0.3) & 4.5(0.7) & 3.0(0.6)\\
			ALESS66.1 & 2.5534(2) & 2.7(0.1) & 1.5(0.3) & 3.1(0.1) & 2.2(0.2) & 3.9(0.2) & 3.2(0.5) & 1.3(0.3)\\
			ALESS67.1 & 2.1228(6) & 3.7(0.2) & 2.6(0.4) & 1.7(0.1) & 1.8(0.3) & 4.9(0.2) & 5.7(0.5) & 3.4(0.7)\\
			ALESS75.1 & 2.5468(3) & 2.6(0.2) & 3.4(0.5) & 1.7(0.2) & 1.9(0.2) & 4.0(0.2) & 4.0(0.6) & 2.1(0.3)\\
			AS2UDS292.0 & 2.1822(1) & 3.1(0.4) & 1.0(0.2) & 3.1(0.4) & 2.0(0.7) & 4.3(0.2) & 3.9(0.3) & 1.8(0.5)\\
			AS2UDS412.0 & 2.5217(8) & 3.7(0.2) & 0.5(0.2) & 1.5(0.1) & 1.3(0.3) & 2.7(0.3) & 1.5(0.8) & 1.1(0.5)\\
		\hline
	\end{tabular}
\tablefoot{Uncertainties are given in the parentheses}
\tablefoottext{a}{Total 870\,$\mu$m continuum flux density estimated from Gaussian fits in the visibility domain.}
\tablefoottext{b}{Total H$\alpha$ flux density.}
\tablefoottext{c}{Half-light radius in major axis at 870\,$\mu$m obtained from {\sc casa} fitting in the visibility domain.}
\tablefoottext{d}{Half-light radius at 870\,$\mu$m obtained from the curve-of-growth analyses based on the ALMA continuum imaging.}
\tablefoottext{e}{H$\alpha$ half-light radius in major axis obtained using {\sc galfit} based on the 2D emission line intensity maps.}
\tablefoottext{f}{H$\alpha$ half-light radius obtained from the curve-of-growth analyses based on the IFU cubes.}
\tablefoottext{g}{H$\alpha$ over 870\,$\mu$m continuum size ratio in which the sizes from the curve-of-growth method are adopted.}
\end{center}
\end{table*}

\section{Analyses and measurements}\label{sec:analyses}
\subsection{SINFONI and KMOS IFU}
For spectral analyses we use the code detailed in \citet{Chen:2017aa} to fit both the emission lines and the near-infrared (NIR) continuum. In short, the code first performs fits to each spectrum with various models including continuum and different combinations of the H$\alpha$, [N\,{\sc ii}], and [S\,{\sc ii}] lines. It then selects the best model based on the Akaike information criterion, in particular the version that is corrected for the finite sample size (AICc; \citealt{Hurvich:1989aa}). Finally, a model fit with line components is considered significant if the fit, compared to a simple continuum-only model, has a lower AICc and provides a $\chi^2$ improvement of $\Delta\chi^2 > 16$\footnote{Equivalent to an S/N $> 4$\,$\sigma$ assuming Gaussian noise and that the noise is not correlated among wavelength channels} for an H$\alpha$-only model, and an additional improvement of $\Delta\chi^2 > 9$ for each additional line. The fits are weighted against the sky spectrum provided by \citet{Rousselot:2000aa} and when calculating $\chi^2$ the wavelength ranges corresponding to the skylines are masked. The velocity dispersion is corrected in quadrature for instrumental broadening. The errors are derived using Monte Carlo simulations; We create fake spectra by injecting the model profile into spectra extracted from randomly selected regions of the data cube with the same circular aperture used for the detection spectrum. The errors are then obtained from the standard deviations between the fit results and the input model.

To measure the total line flux densities, we employ the curve-of-growth analyses, in which the integrated line flux densities are measured with increasing size of circular apertures. The total line flux densities are then obtained at a certain radius beyond which  the line flux densities do not significantly increase, so hitting a plateau. This approach is independent of any 2D surface brightness models and also adopted in some recent work in the literature (e.g., \citealt{Chen:2017aa,Forster-Schreiber:2018aa}). To do such analyses, first the centroid of the circular aperture needs to be determined. We define the centers of H$\alpha$ curve-of-growth analyses to be the centroids of the ellipses best fit to the isophotes of the two dimensional (2D) emission line maps, {specifically the isophotes that have their sizes matched to the corresponding spatial resolution.} The 2D emission line maps are created by performing line fitting on spectra extracted from each individual pixels, with an adaptive binning approach up to 5$\times$5 pixels depending on the signal-to-noise ratios (e.g., \citealt{Swinbank:2006aa,Chen:2017aa}). In pixels where the fitting still fails after 5 × 5 binning to give an adequate S/N, we leave the pixel blank without a fit. The caveat of this approach is that the signals are weighted toward the higher S/N pixels. The outcome of the curve-of-growth analyses is plotted in \autoref{fig2}, the measurements are given in \autoref{table2}, and the 2D emission line maps are shown in \autoref{fig1}.

The H$\alpha$ sizes are measured in two ways. First is to measure the sizes in the curve-of-growth analyses shown in \autoref{fig2}. We use a spline linear fit to the measurements and derive the radius, along with the uncertainties, that corresponds to half of the normalized flux density. The same procedure is applied on the PSFs, and by subtracting the PSF sizes from the measured sizes in quadrature we derive the deconvolved, intrinsic sizes. We apply the same analyses on both H$\alpha$ and the 870\,$\mu$m continuum. The results are given in \autoref{table2}. 

The second method is to use {\sc galfit} (v3.0.5; \citealt{Peng:2010aa}) to conduct single S\'{e}rsic profile fits. The basic procedure closely follows \citet{Chen:2015aa}. Before fitting, we first convert the 2D H$\alpha$ emission line maps to the units in counts, instead of flux density, which is recommended in the {\sc galfit} user manual. The PSF images used for {\sc galfit} are sky-subtracted and properly centered at the peak of the PSFs, and we confirm that all PSFs are Nyquist sampled (FWHM > 2 pixels). We limit the S\'{e}rsic index range to between 0.1 and 4, and leave the rest of the parameters free without constraints. The results are given in \autoref{table2} and they are consistent with those obtained from the curve-of-growth methods. To check whether out results are dependent on the line fitting code, we also perform the S\'{e}rsic fits on the narrow-band H$\alpha$ imaging, which is produced by averaging the wavelength channels within the FWHM of the line, and we again find consistent results. 

{The size of the H$\alpha$ emission has a range of 1-7\,kpc from the curve-of-growth method with a median size of $r_{\text{e,H$\alpha$,cog}} = 3.9\pm0.3$\,kpc (bootstrapped uncertainty), which is consistent with the one in major axis obtained from the S\'{e}rsic profile fits.}

\subsection{ALMA 870\,$\mu$m continuum}
All six SMGs are significantly detected (peak S/N > 5) in both sets (natural weighting and tapered) of the ALMA images (\autoref{fig1}). We measure their 870\,$\mu$m continuum flux densities and sizes in the visibility domain using the {\sc casa} package {\sc uvmodelfit} assuming elliptical Gaussian profiles, except for ALESS75.1, in which there is another serendipitous detection in the map (Hodge et al. 2013; ALESS75.2 is $\sim$10$''$ away so not visible in \autoref{fig1}) so the package {\sc uvmultifit} \citep{Marti-Vidal:2014aa} was used. We have attempted to adopt disk models\footnote{The disk model in {\sc casa} is not an exponential disk but an uniformly bright disk.} in the fitting but in all cases Gaussian models produce better fits with lower $\chi^2$. The results are given in \autoref{table2} along with those derived from the curve-of-growth analyses. 

{The median size derived from the one dimensional curve-of-growth method  ($r_{\text{e,870,cog}}$) is $1.9\pm0.2$\,kpc, which is consistent within errors with the one in major axis based on Gaussian fits in the {\it uv}-plane. The slightly larger sizes in major axis over the one from curve-of-growth (median\{$r_{\text{e,maj,870,uv}}/r_{\text{e,870,cog}}$\}$=1.1$) are consistent with the fact that the FIR continuum is not circular in morphology but elongated (\autoref{fig1}).}

To assess whether there is any missing flux that is resolved out in the higher-resolution maps we measure the total fluxes using the {\sc casa} package {\sc imfit} again assuming Gaussian profiles and compare the results to the total fluxes measured in visibility. We find that the flux ratios of all six SMGs are consistent with unity to within 1\,$\sigma$ with a mean ratio of 1.1$\pm$0.1. We therefore conclude that no significant flux is missing at $\sim0\farcs2$ spatial resolution to within 10\%, consistent with the findings of \citet{Hodge:2016aa}.

\section{Discussions}\label{sec:results}
\subsection{Size measurements}
In \autoref{sec:analyses} we present sizes for both the FIR continuum and H$\alpha$. Now we compare our results to those in the literature. Note that while we show that all methods of deriving sizes reach consistent results, for uniformity from now on we adopt the values obtained from the curve-of-growth method, which is model independent and allows consistent measurements on both data sets. 

Our size measurements of the observed 850\,$\mu$m continuum are in agreement with those reported by \citet{Hodge:2016aa}, who measured sizes on a larger sample of ALESS SMGs, as well as \citet{Simpson:2015aa}, who measured sizes on a sub-sample of the brighter SMGs in the AS2UDS sample. The finding of sizes between 1-2\,kpc is also consistent with studies in other samples of dusty galaxies at $z>1$ (e.g., \citealt{Ikarashi:2015aa,Spilker:2015aa,Tadaki:2017aa,Fujimoto:2018aa}), regardless of them being on the stellar mass - star-formation rate main sequence or not.

The median H$\alpha$ size of our SMG sample, $3.9\pm0.3$\,kpc, is consistent with other SMG samples (e.g., \citealt{Alaghband-Zadeh:2012aa}) but significantly larger than those reported based on other star-forming galaxy samples at $z\sim2$, such as the H$\alpha$ emitters (e.g., median $2.4\pm0.1$\,kpc by \citet{Molina:2017aa} or optical/NIR continuum selected samples (e.g., median 2.5\,kpc by \citet{Forster-Schreiber:2018aa}. However since on average the size of H$\alpha$ agrees with that of optical continuum \citep{Forster-Schreiber:2018aa} and the size of optical continuum positively correlates with the stellar mass (e.g., \citealt{van-der-Wel:2014aa}), this discrepancy can be explained by the fact that our sample SMGs are more massive ($\sim10^{11}$\,M$_\odot$) than that of the other galaxy samples ($\sim10^{10}$\,M$_\odot$).


    \begin{figure}[thb]
    	\centering
    	\includegraphics[scale=0.89]{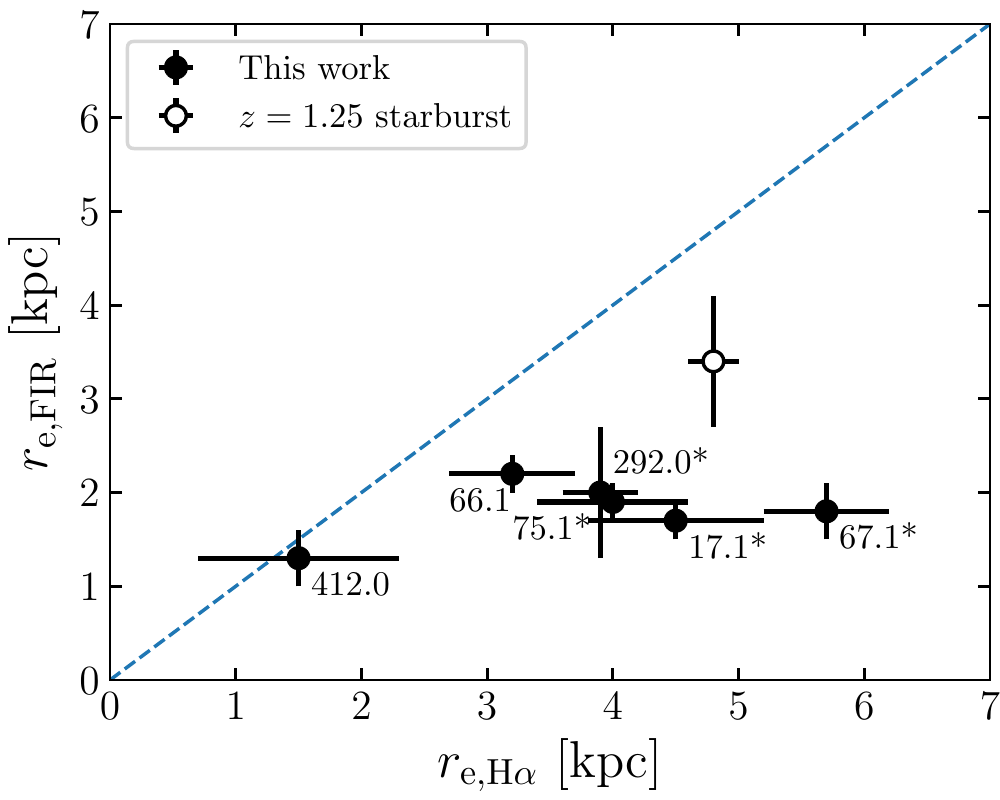}
    	\caption{A plot showing the comparison between the sizes of H$\alpha$ and those of the rest-frame FIR continuum. Our sample SMGs are plotted in filled symbols {along with their IDs, which have their field names removed for clarity. Sources that are identified to host an AGN are marked with an asterisk after the ID.} We also plot in an empty symbol the measurements of a starburst galaxy at $z=1.25$ recently presented by \citet{Nelson:2019aa}. Based on our sample SMGs we find on a median H$\alpha$-to-FIR size ratio of ${2.0\pm0.4}$ with a bootstrapped uncertainty.}
    	\label{fig3}
    \end{figure}
    
The sizes of FIR continuum and H$\alpha$ of our sample SMGs, as well as one dusty galaxy at $z=1.25$ that also has size measurements in both FIR continuum and H$\alpha$\citep{Nelson:2019aa}, are compared in \autoref{fig3}, in which a clear size difference is shown. In all cases H$\alpha$ emission appears larger than the FIR continuum (half differ by $>3\sigma$), with a maximum H$\alpha$-to-FIR size ratio of over three. On average, the size difference in our sample is a median factor of ${2.0\pm0.4}$ with a bootstrapped uncertainty. 

Interestingly we find that SMGs hosting AGN tend to have larger size ratios, {prompting the question of whether AGN is driving the larger H$\alpha$-to-FIR size ratios. For the three X-ray AGN we can infer the expected H$\alpha$ luminosities contributed by the AGN by adopting the $L_{\rm X}-L_{\rm H\alpha}$ correlation deduced by \citet{Ho:2001aa} based on samples of nearby AGN, and we find that $\lesssim$10\% of the measured H$\alpha$ luminosities are contributed by AGN. In addition, the low [N{\sc ii}]-to-H$\alpha$ ratios in all of our sample SMGs, in particular the outskirts, suggest that the ionizing conditions are consistent with those of H{\sc ii} regions, and that H$\alpha$ can be mainly attributed to star formation and not AGN. Finally, there is currently no evidence, including our sample, suggesting that the FIR continuum sizes depend on the presence of an AGN (e.g., \citealt{Harrison:2016ab}). 
	
On the other hand, recently a similar study by \citet{Scholtz:2020aa} on a sample of eight X-ray selected AGN at $z\sim2$ with strong H$\alpha$ and [O{\sc iii}] lines also found a similar H$\alpha$-to-FIR size ratio of $2.3\pm0.3$. This evidence could suggest that somehow AGN is driving the larger H$\alpha$ sizes, or it could be that this is a general feature of FIR luminous galaxies at $z\sim2$ and the current studies are  biased toward galaxies hosting AGN because of their brightness of strong optical lines. Nevertheless, a systematic study on a larger sample, especially including more FIR luminous sources without hosting AGN, is clearly needed to investigate this further.}

The larger H$\alpha$ size over FIR continuum in $z\sim2$ dusty star-forming galaxies is somewhat expected according to recent studies in the literature comparing either FIR and rest-frame optical continuum or H$\alpha$ and rest-frame optical continuum. For example, \citet{Hodge:2016aa} found that the rest-frame optical continuum of their sample of ALESS SMGs is about three times larger than the FIR continuum. This size difference of a factor of 2-3 is almost universally observed in both other similarly FIR-luminous galaxies (e.g., \citealt{Barro:2016aa,Elbaz:2018aa}) and less FIR-luminous star-forming galaxies at $z\sim2$ (e.g.,  \citealt{Tadaki:2017aa,Fujimoto:2018aa}). 

On the other hand, using AO-aided SINFONI IFU data that are matched in spatial resolution to the {\it HST} imaging, \citet{Forster-Schreiber:2018aa} show that on average the H$\alpha$ size of their $z\sim2$ massive star-forming galaxies is identical to the rest-frame optical continuum. \citet{Chen:2017aa} found a similar result on one SMG, ALESS67.1, which is included in our sample.  Four of our sample SMGs have {\it HST} {\it H}-band imaging, and the results of the curve-of-growth analyses are plotted in \autoref{fig2}. On average we find a H$\alpha$-to-optical size ratio of 1.0$\pm$0.1. 

All these results suggest that on average H$\alpha$ is similar in size to the rest-frame optical continuum, and both are a factor of 2-3 larger than the FIR continuum. The much smaller size of FIR continuum compared to almost any other tracers, including molecular gas (e.g., \citealt{Ginolfi:2017aa,Chen:2017aa,Calistro-Rivera:2018aa}), challenges the typical assumption regarding the treatment of dust attenuation in the SED modeling (e.g., \citealt{Simpson:2017ab}). In the following sections we investigate and discuss in detail some possible implications.

\subsection{Star formation rates}
There are a few possible implications of the size difference between FIR continuum, H$\alpha$ and optical-to-infrared (OIR) continuum. We start with the measurements of star-formation rates. Note the UV-to-NIR photometry of ALESS66.1 is contaminated by a nearby quasar at $z=1.31$ \citep{Simpson:2014aa,Danielson:2017aa}. We therefore exclude it from the analyses from now on.

\subsubsection{Dust correction of H$\alpha$}\label{sec:ha}
	
Under the assumption that the total star-formation rates can be estimated through either attenuation-corrected H$\alpha$ or UV-to-FIR continuum photometry, one possible implication of the size difference has to do with the attenuation correction of H$\alpha$-based SFRs, particularly in cases where FIR measurements are not available. To account for dust attenuation a typical approach is to assume a foreground dust screen, which is co-spatial in the projected sky with the underlying UV/optical star-formation tracers. {This is one of the fundamental assumptions of some of the popular SED modeling involving corrections of dust attenuation, including {\sc hyperz} \citep{Bolzonella:2000aa} and {\sc eazy} \citep{Brammer:2008aa}.} However with spatial mismatches between H$\alpha$ and FIR continuum these assumptions need to be further examined and the possible impact needs to be understood. 

To understand such an impact, if any, we first aim to compare the attenuation-corrected H$\alpha$-based SFRs and the dusty SFRs derived from the total infrared luminosities. That is, instead of adopting the energy-balance approach we model the UV-to-NIR and MIR-to-radio SEDs separately, so mimicking the traditional approach of deriving total SFRs without FIR measurements and then use the FIR-inferred SFRs to validate this approach. In principle, a significant size difference between dust and H{\sc ii} regions may result in the attenuation-corrected H$\alpha$-based SFRs being systematically lower than the dusty SFRs, or both measurements being uncorrelated, or both.

To obtain the attenuation-corrected H$\alpha$ SFRs, the best approach is to also measure the H$\beta$ luminosity and estimate the nebular attenuation through the Balmer decrement (e.g., \citealt{Reddy:2015aa}). However H$\beta$ is not available in all but a marginal detection from one SMG, ALESS75.1. To be able to apply a consistent methodology across the sample, alternatively we opt to use the stellar attenuation derived from the UV-to-NIR SED fitting, further corrections on top of the stellar attenuation may need to be applied (e.g., \citealt{Calzetti:2000aa,Wuyts:2013aa,Price:2014aa}).

To model the UV-to-NIR SEDs we use {\sc hyperz} \citep{Bolzonella:2000aa}, a $\chi^2$ minimization code to fit a set of model SEDs on the observed photometry. The model SEDs are based on synthetic SED templates whose intrinsic shape is characterized by the star-formation history (SFH). The synthetic SEDs are further modified according to dust reddening, Lyman forest, and redshifts. The four synthetic SED templates considered are created with spectral templates of \citet{Bruzual:2003aa}, assuming solar metallicities, with different SFHs: a single burst (B),  two exponentially decaying SFHs with timescales of 1 Gyr (E) and 5 Gyr (Sb), and constant star formation (C). The undetected photometric measurements are set to a flux of zero with uncertainties equal to 1\,$\sigma$ of the limiting magnitude of that filter. We follow the \citet{Calzetti:2000aa} law to allow total attenuation ($A_V$) between 0 to 5 in steps of 0.01. The age of the galaxy must be younger than the age of the Universe. 


Our methodology is similar to that of \citet{Simpson:2014aa}, who also conducted {\sc hyperz} fitting on ALESS SMGs. However they did not have full spectroscopic redshift information. Therefore as a check we first compare our fitting results on the ALESS SMGs against those of \citet{Simpson:2014aa}, by allowing the redshift to vary between 0 and 6 in steps of 0.1. We confirm that we are able to reproduce their results. We then determine $A_V$ by setting the redshift range to the spectroscopic redshifts with uncertainties measured from H$\alpha$. We find that under one synthetic SED template $A_V$ are always invariant within the spectroscopic redshift uncertainties (we adopt $\pm$3\,$\sigma$). However the variations become significant from template to template, namely they are affected by the SFHs adopted. We therefore perform fits by considering one SED template at a time and take all the $A_V$ values from fits that have $\Delta\chi^2<1$ from the lowest $\chi^2$. When deriving the dust-corrected SFRs we propagate the range of these $A_V$ values into the uncertainties. We find typical $A_V$ values of 1 to 3 for SED sampled down to the rest-frame UV, consistent with previous findings of SMGs \citep{Takata:2006aa,Wardlow:2011qy,Simpson:2014aa,da-Cunha:2015aa}.

	\begin{figure}[!htb]
		\centering
		\includegraphics[scale=0.9]{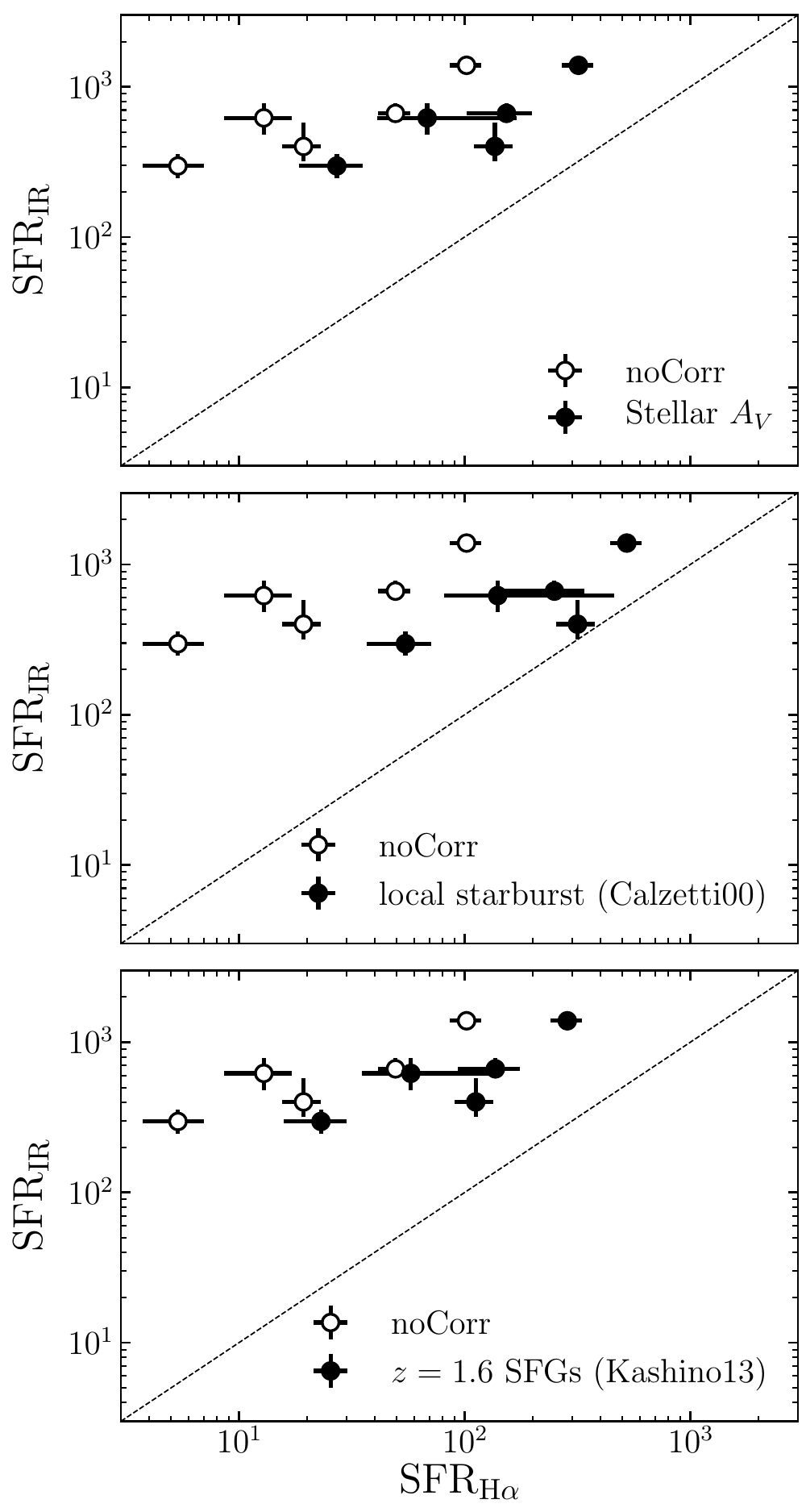}
		\caption{The comparison between the IR-based SFRs and the H$\alpha$-based SFRs, except for ALESS66.1, where the UV-to-NIR photometry is contaminated by a foreground nearby quasar. The derivations of SFRs are described in detail in \autoref{sec:ha}. In each panel the empty symbols are plotted based on the measured H$\alpha$-based SFRs without attenuation correction (noCorr), and the filled symbols are plotted based on the H$\alpha$-based SFRs corrected for attenuation using various methods; In the top panel we adopt the stellar A$_V$ derived from the SED fitting, and in the middle and bottom panel we adopt the attenuation of the H{\sc ii} regions based on a fractional correction to the stellar A$_V$ provided by \citet{Calzetti:2000aa} and \citet{Kashino:2013aa}, respectively. While the exact amount of correction for attenuation is still under debate, we find that even if we adopt the largest correction provided by \citet{Calzetti:2000aa}, the H$\alpha$-based SFRs are still on average a factor of $3\pm1$ lower than the IR-based SFRs.}
		\label{fig4}
	\end{figure}
	
We now move on to the fitting of the FIR SEDs. We adopt the approach of template fitting. In particular we use the library of 185 template SEDs constructed by \citet{Swinbank:2014aa}, who included local galaxy templates from \citet{Chary:2001aa}, \citet{Rieke:2009aa} and \citet{Draine:2007aa}, as well as the SEDs of the well-studied high-redshift starbursts SMM J2135−0102 ($z = 2.32$) and GN20 ($z = 4.05$) from \citet{Ivison:2010aa} and \citet{Carilli:2011aa}, respectively. The MIR-to-radio photometry of the ALESS SMGs, including MIPS 24, PACS 70 and 160\,$\mu$m, SPIRE 250, 350, and 500\,$\mu$m, ALMA 870\,$\mu$m, and VLA 1.4\,GHz, are derived by \citet{Swinbank:2014aa}. The same methodology is applied to the AS2UDS SMGs \citep{Stach:2019aa}. We fit the MIR-to-radio photometry using $\chi^2$ minimization and adopt the spectroscopic redshifts measured by our IFU observations. We then derive the infrared luminosity by integrating the best-fit template SED, as well as all acceptable SEDs based on their $\chi^2$ values, over the rest-frame 8-1000\,$\mu$m range.

Having the fitting results in hand, we estimate the star-formation rates based on H$\alpha$ and infrared luminosities by adopting the calibration provided by \citet{Kennicutt:2012aa} \footnote{{The conversion from infrared luminosities to SFRs heavily depends on the star-formation history, varying by a factor of around five given different star formation timescales \citep{Calzetti:2013aa}. Ideally one should adopt the conversion based on the best-fit SED template from {\sc hyperz}. However as shown in previous studies of SMGs, the UV-to-NIR continuum tends to spatially offset from the FIR continuum \citep{Chen:2015aa,Hodge:2016aa}. Therefore the star-formation history derived from UV-to-NIR photometry likely does not reflect the true star-formation history in the dust regions emitting FIR continuum. The conversion from \citet{Kennicutt:2012aa} represents roughly the mean value of the possible range provided by \citet{Calzetti:2013aa}. We therefore expect the uncertainty of the conversion due to unknown star-formation history contributes mostly to the scatter of the correlation, not the normalization.}}. We plot the results in \autoref{fig4}. First we find that the H$\alpha$-based SFRs without attenuation correction are on average a factor of $21\pm15$ in median (with bootstrap error) lower than infrared-based SFRs. Similar discrepancies have been reported in other SMG samples \citep{Swinbank:2004aa, Casey:2017aa}. We then correct the H$\alpha$ luminosities using the stellar attenuation via $L_{\rm H\alpha} = L_{\rm H\alpha,obs}\times10^{0.4A_{V,star}}$, in which $A_{V,star}$ is obtained through the UV-to-NIR {\sc hyperz} fitting. After correcting for dust attenuation, the infrared SFRs are still systematically higher, but now by a factor of $4\pm2$. 

The systematically lower H$\alpha$-based SFRs after correcting for stellar attenuation suggests that, given the assumption that the total SFRs can be obtained either from infrared luminosity or attenuation-corrected H$\alpha$, an additional correction on top of the stellar attenuation is needed for H$\alpha$. On the other hand, if the discrepancy persists after further corrections, the aforementioned assumption may be challenged and it could suggest that the bulk of the dusty star formation {may not be traced through rest-frame optically detectable regions including H{\sc ii} regions, as partly supported by our findings of size difference.}


 

To assess the amount of further correction, we now look at the possible differences between nebular and stellar attenuation. Using the Balmer decrement, the comparison of the two has been extensively studied, in both local samples \citep{Calzetti:2000aa,Wild:2011aa,Kreckel:2013aa} and higher redshifts \citep{Kashino:2013aa,Price:2014aa,Reddy:2015aa,Theios:2018aa}. In particular, the seminal work of \citet{Calzetti:2000aa} found that using a sample of local starbursting galaxies the nebular attenuation is larger with a relation of $E(B-V)_{star}=0.44E(B-V)_{neb}$. However a variety of results have been found in other samples. While the consensus has yet to be reached since the relation likely depends on galaxy properties \citep{Wild:2011aa,Price:2014aa,Puglisi:2016aa}, there is a growing evidence showing that at $z>1$, between the nebular and stellar attenuation, the discrepancy appears to be smaller but the correlation is scattered \citep{Kashino:2013aa,Reddy:2015aa,Puglisi:2016aa,Theios:2018aa}.

For this exercise, we adopt two representative results from \citet{Calzetti:2000aa} and \citet{Kashino:2013aa}, as the former represents the largest discrepancy between nebular and stellar attenuation found so far, and the later presents a galaxy sample that has properties closer to those of our sample in redshift, stellar mass, and SFR. Calzetti et al. found $E(B-V)_{star}=0.44\pm0.03E(B-V)_{neb}$ and Kashino et al. found $E(B-V)_{star}=0.70\pm0.08E(B-V)_{neb}$\footnote{The original relation found by Kashino et al. is $E(B-V)_{star}=0.83\pm0.1E(B-V)_{neb}$, which explicitly assumes a Calzetti attenuation curve on both Balmer decrement and stellar continuum. However \citet{Calzetti:2000aa} adopted the Calzetti attenuation curve for the stellar continuum but a Milky Way attenuation curve on H$\alpha$ \citep{Cardelli:1989aa}. We therefore convert the Kashino et al. result based on the same assumption of the attenuation curves used in \citet{Calzetti:2000aa}.}



Given $A_\lambda = \kappa(\lambda)E(B-V)$ in which $\kappa(\lambda)$ is the attenuation curve\footnote{$\kappa(H\alpha;6565\AA) = 2.54$ assuming Milky Way attenuation curve \citep{Cardelli:1989aa} and $\kappa(V;5530\AA)=4.02$ assuming Calzetti attenuation curve.}, we deduce a total attenuation relation of $A_{H\alpha} = 1.4\pm0.1A_{V,star}$ and $A_{H\alpha} = 0.9\pm0.1A_{V,star}$ based on the result of \citet{Calzetti:2000aa} and \citet{Kashino:2013aa}, respectively. We apply the relation to our measurements and plot the results in \autoref{fig4}.

Despite applying further corrections, the attenuation-corrected H$\alpha$-based SFRs still cannot account for all the SFRs revealed in the infrared, missing at least a median factor of $3\pm1$ considering the most aggressive correction based on the Calzetti attenuation curve. The Spearman correlation coefficient is 0.6 ($p=0.2$) so the two SFRs appear possibly correlated among these five galaxies. However given the discrepancy between the two SFRs even after accounting for attenuation for H$\alpha$, the correlation could be driven by the global properties of the galaxy such as gas fraction or the dynamical environment, instead of them tracing the same part of star-forming regions, which is partially supported by our findings of size difference between the two SFR tracers. {It is also possible that dust is spatially mixed with H{\sc ii} regions, instead of acting as a foreground screen, and that dust column density reaches a level that H$\alpha$ becomes optically thick, in particular in the central regions, so H$\alpha$ does not reflect the full attenuation. Evidence of this possibility has been shown in studies of local (Ultra) Luminous Infrared Galaxies ((U)LIRGs), where the attenuation derived from Pa$\alpha$/Br$\gamma$ or Br$\gamma$/Br$\delta$ in near-infrared is slightly higher than that derived from Balmer decrement (e.g., \citealt{Piqueras-Lopez:2013aa}). Interestingly, for the local ULIRGs there is also evidence showing that their IR-based SFRs are higher than the attenuation-corrected, H$\alpha$-based SFRs, by a factor of 2-30 \citep{Garcia-Marin:2009aa}. In the next section we further investigate the reasons driving the mismatching SFRs, in particular the size difference between FIR continuum and H$\alpha$.}



\subsubsection{Total star-formation rates}\label{sec:sfr}
In \autoref{sec:ha} we discussed the implications due to the spatial mismatches between the FIR continuum, H$\alpha$ emission, and optical continuum. In particular when considering SFRs, the attenuation corrected H$\alpha$-based SFRs are systematically lower than the IR-based SFRs. Given the total SFRs at high redshifts are frequently either derived from IR+UV\footnote{In our case UV-based SFRs are $\sim$1\% of IR-based SFRs so negligible.} or attenuation corrected H$\alpha$ (e.g., \citealt{Shivaei:2016aa}), the discrepancy merits further discussions. We explore two closely related issues: One on the attenuation correction, and the other about the spatial mismatches between the bulk of dust and the H{\sc ii} regions.

	\begin{figure*}[thb]
		\centering
		\includegraphics[scale=0.73]{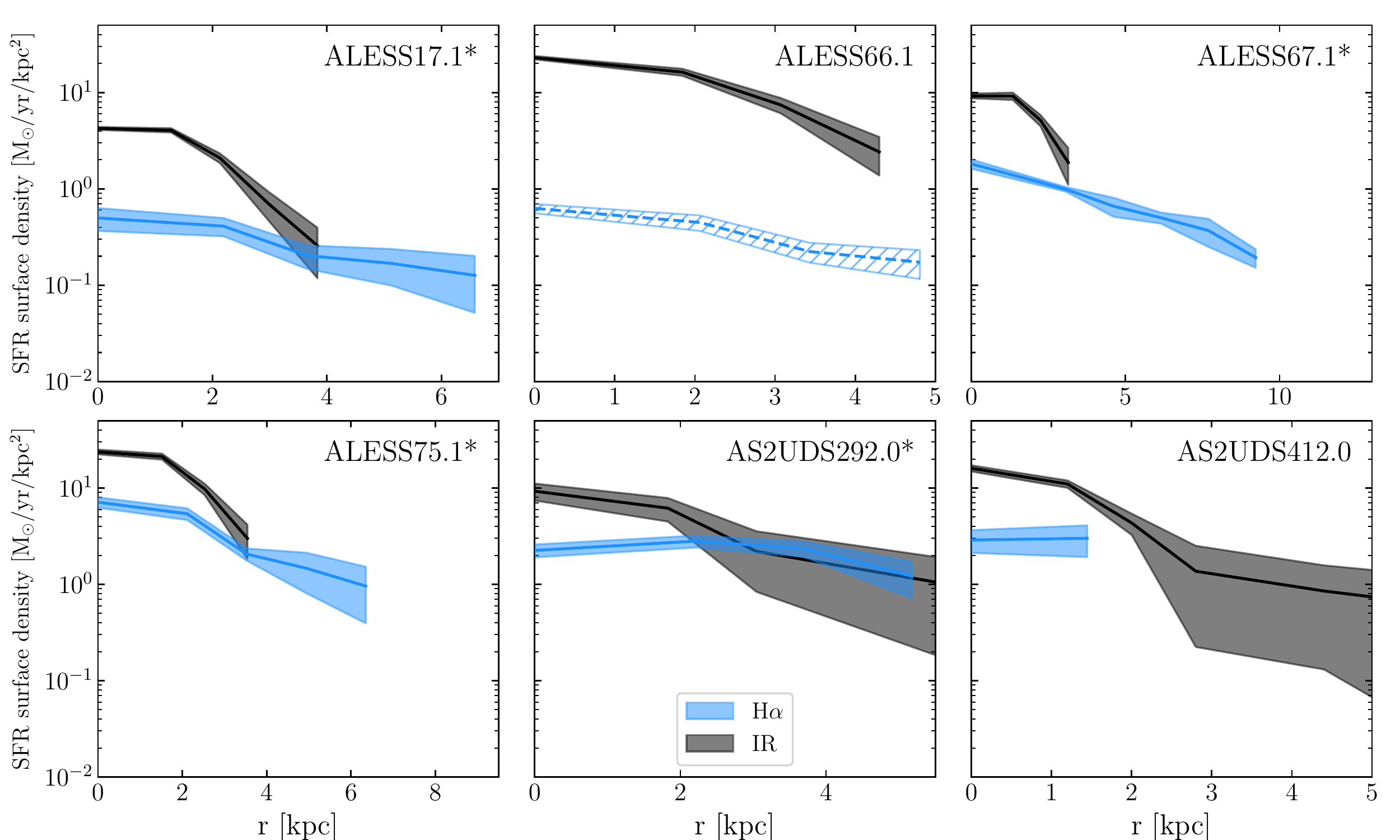}
		\caption{The SFR density profiles based on the curve-of-growth analyses (\autoref{fig2}), which are deconvolved according to the PSF. Sources that are identified to host an AGN are marked with an asterisk after the ID. For the IR-based profiles we assume that they follow the morphology of the observed 870\,$\mu$m emissions, meaning the IR-based SFRs in each radial bin is the fraction of the total 870\,$\mu$m flux times the total SFRs. The H$\alpha$-based profiles are derived by differentiating the curve-of-growth results, converting the H$\alpha$ luminosity to SFRs according to Kennicutt \& Evans (2012), and adopting an attenuation correction of \citet{Calzetti:2000aa}, except for ALESS66.1, of which the OIR photometry is contaminated by a foreground quasar (\autoref{fig1}). For ALESS66.1 we plot the profile without correcting for attenuation. For clarity we only plot the bins that have $\ge$1\,$\sigma$ measurements.}
		\label{fig5}
	\end{figure*}

On the attenuation correction, in order to align both SFRs, the correction needs to be about 50\% more on top of the extra correction provided by \citet{Calzetti:2000aa}, which is similar to what was found by \citet{Wuyts:2013aa}. However such an aggressive correction based on the Balmer decrement has never been observed (e.g., \citealt{Price:2014aa}). In fact, studies of $z\sim2$ star-forming galaxies have found the opposite with significantly smaller values (e.g., \citealt{Kashino:2013aa,Reddy:2015aa}). We do note that these studies mostly focus on less obscured, lower stellar mass, and lower SFR galaxies in comparison with our sample SMGs. However if the IR-based and H$\alpha$-based SFRs were indeed to be matched, it would have suggested that globally the relation between the nebular $A_V$ and stellar $A_V$ is very different in the SMGs compared to that in the more typical star-forming galaxies. {A more likely situation is that the bulk of the obscured star formation is not traced by either the H$\alpha$ or stellar emissions, due to a combination of a few factors including global-scale size difference and very high obscuration in the central dusty regions.} In the following we discuss further these two possibilities.

\begin{figure*}[!htb]
	\centering
	\includegraphics[scale=1.05]{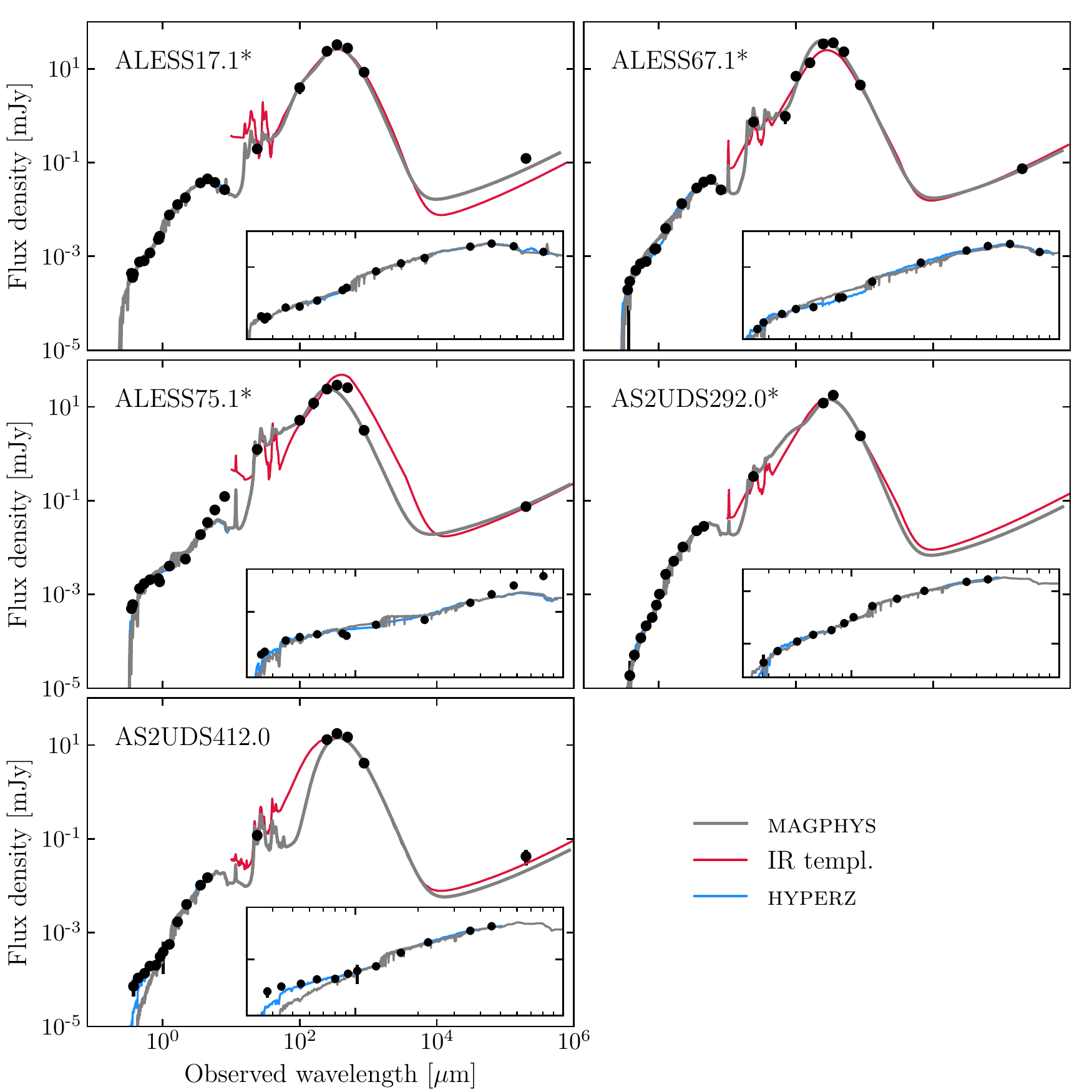}
	\caption{UV-to-radio SEDs for each of our sample SMGs except ALESS66.1, of which the UV-to-NIR photometric measurements are contaminated by a foreground quasar \citet{Simpson:2014aa}. Sources that are identified to host an AGN are marked with an asterisk after the ID. The measurements are plotted as black points and the best-fit models from {\sc magphys}, {\sc hyperz}, and the IR templates compiled by \citet{Swinbank:2014aa}, are plotted as grey, blue, and red curves, respectively. For clarity we do not show the uncertainty however they are discussed in the text, {including the fact that the IR luminosities estimated by the two methods agree with each other within the uncertainty.} The insets are zoom-ins in the OIR regime from 0.3 to 10\,$\mu$m in log-log scale. We find significant differences in higher $L_{\rm IR}$ sources between the best-fit models of {\sc magphys} and {\sc hyperz}, and in all these discrepant cases the {\sc hyperz} modeling provides a better fit with a lower $\chi^2$.}
	\label{fig6}
\end{figure*}

To quantify how much {the global-scale size difference} plays a role, one way is to estimate how much of the H$\alpha$-based SFRs originate from outside the bulk of dust as traced by the FIR continuum. Our data allow a simple one dimensional (1D) assessment. To do so, in \autoref{fig5} we plot the SFR densities as a function of radial distances based on the FIR and H$\alpha$ measurements. Since we do not have spatially resolved information of other FIR bands we assume that the IR-based SFR density profile follows the morphology of the observed 870\,$\mu$m emissions, as derived from the curve-of-growth analyses. That is, the IR-based SFRs in each radial bin is the fraction of the total 870\,$\mu$m flux in that bin multiplied by the total SFRs. This method effectively assumes a constant dust temperature and dust opacity, which is likely not true since evidence of negative temperature gradient (hot to cold from center to outskirts) has been found recently in SMGs \citep{Calistro-Rivera:2018aa}. However a negative gradient of dust temperature would mean an even more compact SFR density distribution. 

For the H$\alpha$ profiles, we simply differentiate the curve-of-growth results, convert the H$\alpha$ luminosities to SFRs according to \citet{Kennicutt:2012aa}, and adopt the stellar $A_V$ with a further correction to the nebular $A_V$ based on \citet{Calzetti:2000aa}, except for ALESS66.1, of which the OIR photometry is contaminated so it is excluded in the following discussions. Note the total attenuation is based on integrated photometry so the correction is the same across all radial bins. While it is expected that the total attenuation has a negative gradient so is higher in the central regions \citep{Wuyts:2012aa,Nelson:2016aa,Liu:2017aa}, the total attenuation derived from the integrated photometry likely reflects the averaged conditions across the whole galaxy. We discuss the consequences of this scenario in detail in the later paragraphs. Finally, to show the intrinsic distributions, all profiles are deconvolved in quadrature according to the PSF.

For the simplest quantification, we find that the fraction of H$\alpha$-based SFRs that are located outside the central dusty regions ranges from zero (AS2UDS412.0) to {40}\% (ALESS67.1). The fraction would decrease if there is a negative gradient of $A_V$, but increase with a negative gradient of dust temperature. {We also note that different centroids are adopted for 870\,$\mu$m continuum and H$\alpha$ in the curve-of-growth analyses (\autoref{fig1}), which has however a negligible impact such that the fraction would only increase by a maximum of 5\% if the same centroids had been adopted.} From this exercise, while it appears that in 1D profile the H$\alpha$-based SFRs are more extended relative to the FIR-based SFRs, the bulk of H$\alpha$-based SFRs spatially overlap with the FIR-based SFRs. {That is, the size difference in the global scale between the FIR continuum and H$\alpha$ is not the dominant factor that drives the mismatching SFRs estimated by these two SFR tracers. The dominant factors are likely linked to heavy dust attenuation in the central regions.} 

\begin{figure*}[thb]
	\centering
	\includegraphics[scale=0.9]{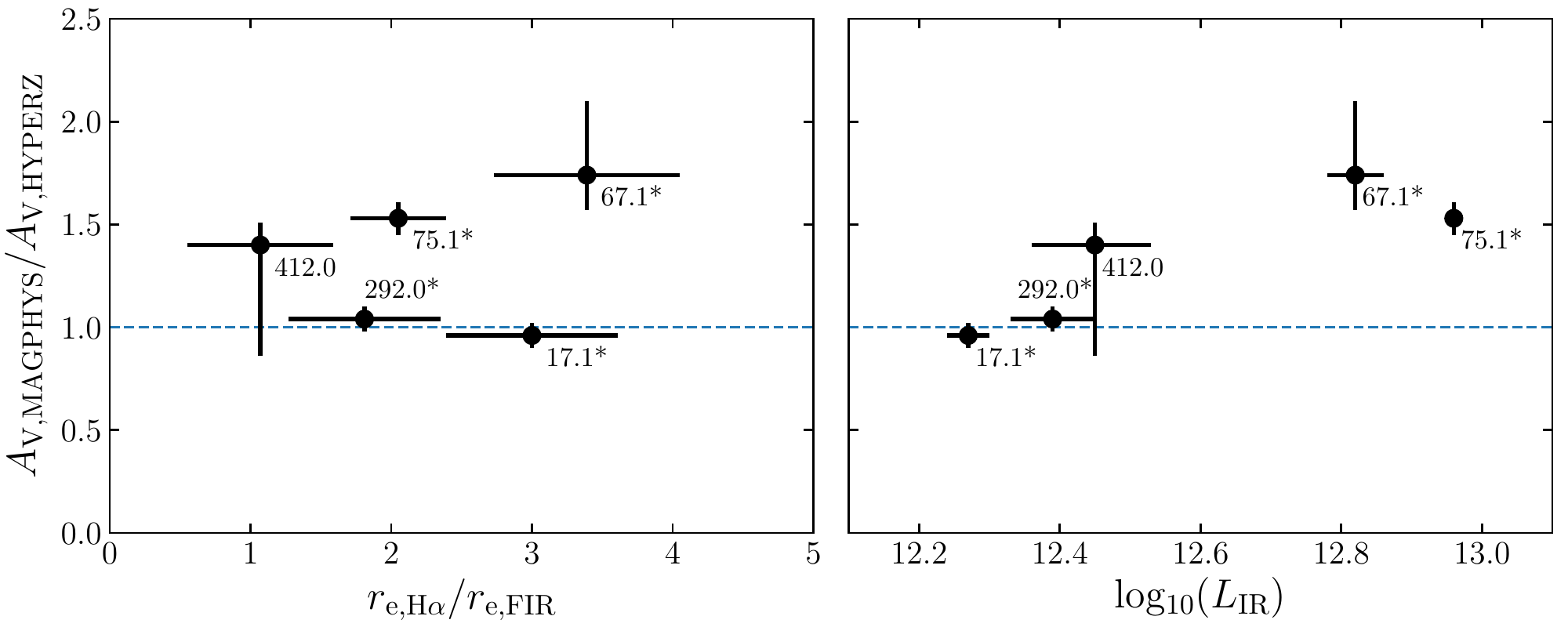}
	\caption{Comparison of the ratio of the total attenuation derived from the best-fit model of {\sc magphys} to that of {\sc hyperz}, the size ratio of H$\alpha$ to the FIR continuum ({\it left}), and the total infrared luminosity ({\it right}). {Data points are plotted along with their corresponding IDs, which have their field names removed for clarity. Sources that are identified to host an AGN are marked with an asterisk after the ID.} We find a correlation between the ratio of attenuation and the infrared luminosity.}
	\label{fig7}
\end{figure*}

Indeed, by estimating the hydrogen column density through dust masses and FIR continuum sizes, \citet{Simpson:2017ab} found that by assuming a foreground screen of dust geometry, the averaged total attenuation in the central dusty regions of SMGs is $A_V=540^{+80}_{-40}$, suggesting that effectively all of the optical-infrared (OIR) emission that is spatially coincident with the far-infrared emission region is completely extinguished by dust. In addition, in the most recent work by \citet{Hodge:2019aa}, at a spatial resolution of $\sim$500\,pc they found that the FIR continuum of SMGs becomes clumpy and structured in shapes of spiral arms, bars, and rings. In such a case the total attenuation of the dusty regions could exceed the values estimated by \citet{Simpson:2017ab}, suggesting that bulk of the dusty star formation as traced by FIR continuum is spatially decoupled from H$\alpha$ on sub-kpc scales. {If we take the median value from \citet{Simpson:2017ab}, which is $A_{\rm V}=540$, and apply the correction to the H$\alpha$-based SFRs, the resulting SFRs would be on the order of 10$^{0.4\times540}\sim10^{216}$\,M$_\odot$ yr$^{-1}$, which is unrealistic. It would still be unrealistically high if we only apply this correction at the central regions where most of the dust is located.} 

On the other hand, another possibility could be that instead of acting as a foreground screen, the dust is mixed with the H{\sc ii} regions. In this case the FIR continuum would not be spatially decoupled from H$\alpha$, but given the high column density H$\alpha$ is no longer optically thin therefore not reflecting the total attenuation. Both scenarios would lead to a significant underestimation of the total SFRs from H$\alpha$. More data on IR continuum along with the Balmer and even the Paschen lines with very high angular resolutions ($<0.1"$), presumably from ALMA and ELT, should shed more light on this issue.

\subsection{SED modeling and dust distribution}
Another possible implication of size difference is the SED modeling, in particular those employing the energy-balance approach. We now discuss this aspect in detail. 

\subsubsection{Comparison to the energy balance approach}\label{sec:magphys}

The spatial mismatch we see between dust and OIR emission may have implications for energy balance SED modeling. In particular, for our sample SMGs, one can imagine that because the majority of dust is not co-located with the OIR emissions, the attenuation estimated from OIR alone may be smaller than the attenuation derived from the energy balance approach. That is, the energy balance fitting may deduce a higher attenuation solution in OIR in order to provide a better fit to the FIR photometry, in particular for high luminosity sources. To test this possibility, we also model our UV-to-radio SEDs using the {\sc magphys} code. We adopt the high-$z$ edition, which includes the modeling in the radio bands, as well as the Lyman absorptions in the rest-frame UV from the intergalactic medium \citep{da-Cunha:2015aa}. We plot the best-fit models in \autoref{fig6}, along with the photometric measurements, and the fitting results of {\sc hyperz} and those using the infrared templates.

We first compare the infrared luminosities estimated from the SED templates\footnote{The total infrared luminosity given by {\sc magphys} is integrated over 3 to 1000\,$\mu$m. We therefore adopt the same wavelength range to compute the total infrared luminosity from the best-fit SED templates.} and those from {\sc magphys}. We confirm that both values are statistically consistent with each other in all sources. We then turn to $A_V$, in which we find significant differences in some sources. As seen in \autoref{fig6}, ALESS67.1, ALESS75.1, and AS2UDS412.0 appear to have different best-fit models in the UV-to-NIR regime, although the fit from {\sc hyperz} shows a better agreement with the data. Interestingly, in these three cases where the $A_V$ differ, the values obtained from {\sc magphys} are all higher. 

Note that given the different assumptions of the attenuation curves in these two codes, ideally the $A_V$ values from the UV-to-NIR regime should also be deduced from {\sc magphys} by turning off the fitting of the MIR-to-radio part. Unfortunately the public version of {\sc magphys} does not support such an option. However, the fact that the $A_V$ values are in excellent agreement with each other in ALESS17.1 and AS2UDS292.0, as expected from a good match shown in \autoref{fig6}, suggests that the systematic difference between the $A_V$ values can be neglected. {We have also tried to homogenize the two codes by running the {\sc hyperz} fitting  using only the $\tau$-decay SFHs similarly adopted by {\sc magphys}, effectively removing the bursting SFH in our adopted method for {\sc hyperz}. We find that the output values vary within uncertainties and the conclusions remain unchanged. Since stellar mass is not one of the direct outputs from {\sc hyperz} we estimate the stellar masses by converting the absolute magnitude, which is one of the outputs from {\sc hyperz}, based on some estimates of mass-to-light ratios. In particular we take the H-band absolute magnitude from {\sc hyperz} and the mass-to-light ratios from \citet{Simpson:2014aa} where the ratios were derived based on the \citet{Bruzual:2003aa} simple stellar population models. We find that all but one (AS2UDS292.0) have their stellar mass estimates from the two codes in agreement with each other within uncertainties. The disagreement on AS2UDS292.0 is mainly caused, as expected, by the fact that the {\sc hyperz} fitting finds the best solution based on the bursting SFH. If we force a $\tau$-decay like fitting then the stellar mass estimated from {\sc hyperz} agrees with that from {\sc magphys}, with a little change in A$_V$ so again the conclusions are not sensitive to this change.}

To further understand what causes the different $A_V$ values, in \autoref{fig7} we first plot the $A_V$ ratios as a function of the size ratios between H$\alpha$ and the FIR continuum. Naively, one may expect that a better agreement in spatial distribution between dust and the OIR emissions would lead to a better agreement in $A_V$. That is, the more the sizes differ the more the $A_V$ differs. Interestingly we do not observe such a trend. Instead, we find a more prominent correlation between the $A_V$ ratios and the total infrared luminosity (\autoref{fig7}). While a larger sample is clearly needed to confirm, this correlation may suggest that indeed the energy balance approach deduces a higher $A_V$ in cases of high infrared luminosity, sacrificing a poorer fit in UV-to-NIR in return for an overall smaller $\chi^2$ value. 

It is understandable that the discrepancy is the largest toward the higher infrared luminosity end, especially when the infrared luminosity is so high that the best-fit UV-to-NIR models cannot account for it. On the other hand, the more interesting question is perhaps why the $A_V$ values agree in the lower infrared luminosity end, despite the spatial mismatches. Based on \citet{da-Cunha:2008aa,da-Cunha:2015aa}, {\sc magphys} adopts a two-component model (birth cloud and diffuse ISM) to describe the attenuation of stellar emission at the UV-to-NIR regime. Since the UV-to-NIR emissions largely originate from the stars in the diffuse ISM, by increasing the attenuation in the birth clouds, it is possible to boost the infrared luminosity without significantly increasing the total attenuation, which is $A_V$. Indeed, for the two sources, ALESS17.1 and AS2UDS292.0, where the modeling of {\sc hyperz} and {\sc magphys} agrees the best, the optical depth seen by the stars in the birth clouds (defined as $\tau_V$ in {\sc magphys}) is a factor of 2-3 higher than the rest of the sources, and the fraction of $\tau_V$ seen by stars in the diffuse ISM (defined as $\mu$ in {\sc magphys}) is a factor of 2-3 smaller. One of the consequences of these ad-hoc tunings is that the infrared emissions only contributed from the birth clouds, such as the PAH and the MIR emissions, could become unrealistically high. Unfortunately this is the regime where we are lacking the data. Future missions targeting MIR such as {\it JWST} and {\it SPICA} will be able to shed more light on this issue.

In short summary, we find that the level of impact due to spatial mismatches on the energy balance approach of SED modeling depends on the physical properties. For example, they have a negligible impact on the total IR luminosity. However on the other hand, for IR luminous ($L_{\rm IR} \gtrsim 10^{12.6-12.8}$\,$L_\odot$) galaxies significant impact can be seen in total attenuation, which is intrinsically related to stellar age, star-formation history, and stellar mass. For less IR luminous galaxies the impact could be more subtle, possibly related to the fractional contribution of total attenuation between the diffuse ISM and the birth clouds.

\subsubsection{Three-component dust distributions}\label{sec:dustdis}
\begin{figure}[thb]
	\centering
	\includegraphics[scale=0.28]{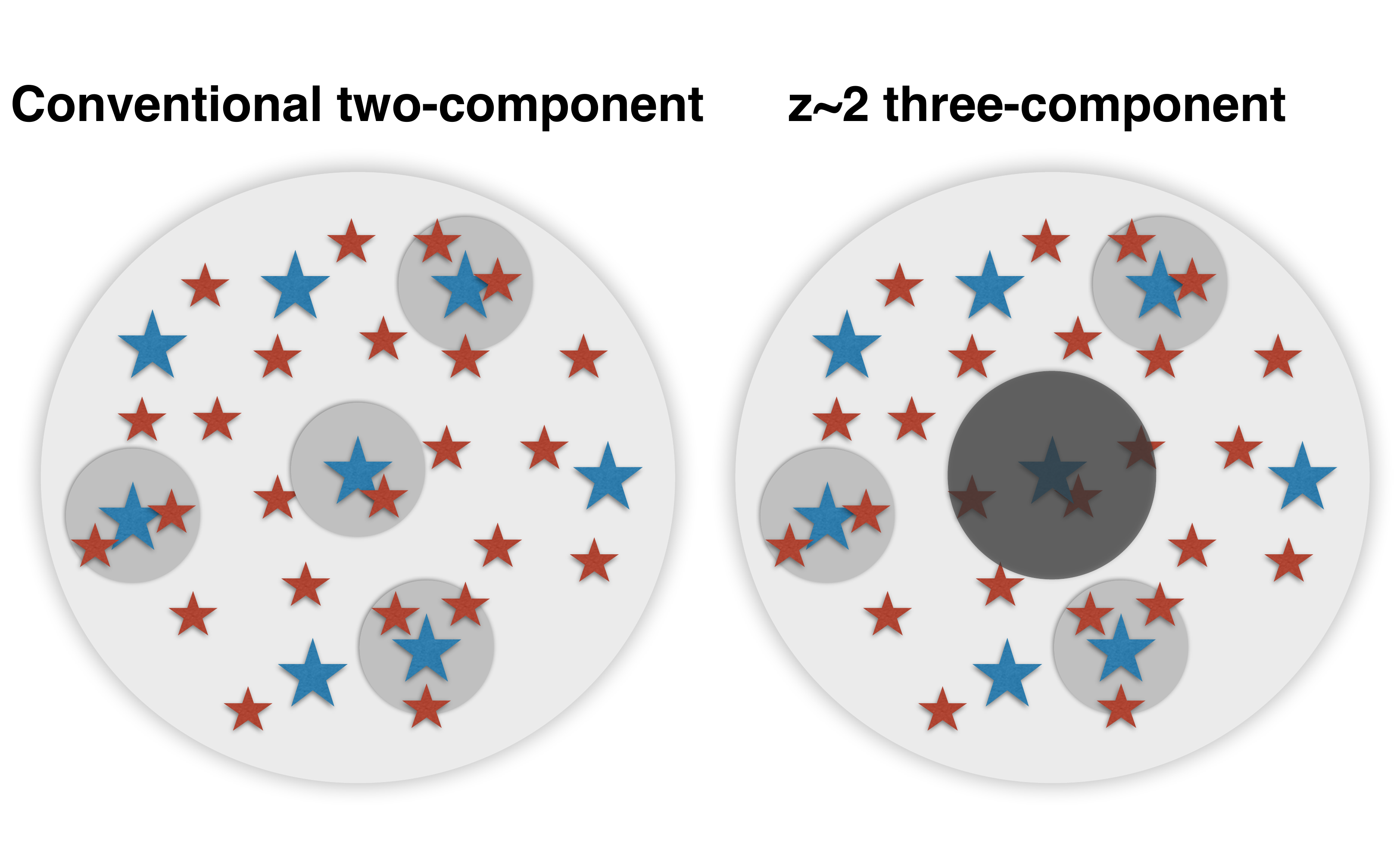}
	\caption{A schematic figure showing the conventional two-component view of the dust distribution and the {postulated} three-component model that is supported by our measurements as well as literature studies on some $z\sim2$ galaxies with lower IR luminosities. The red stars represent older stellar populations that dominate the integrated rest-frame UV to NIR,  while the blue stars represent the young star-forming H{\sc ii} regions traced by H$\alpha$. The grey regions show the rough distributions of dust with the opacity reflecting their total attenuation, meaning the darker regions are more obscured than the lighter ones.}
	\label{fig8}
\end{figure}

As briefly mentioned in the previous section, it is normally perceived that there are in general two main components for dust distribution (or attenuation); the diffuse ISM which encompasses mostly older stars and older star-forming regions revealed by the UV radiation, and the birth clouds tracing mostly H{\sc ii} regions where more intensive and younger star formation occurs (\autoref{fig8}; \citealt{Calzetti:1994aa}). This is the essential assumption some of the SED models adopt, including those employ  the energy balance approach (e.g., {\sc magphys} and {\sc cigale}). The advantage of this model, and some variations based of it, can explain most of the observational results regarding the different attenuation observed between H$\alpha$ and the stellar continuum. For example, the generally higher color excess of H$\alpha$ compared to that of stellar continuum is a natural consequence of this model. In addition, the fractional distribution between the birth clouds and the diffuse ISM can explain some correlations between the galaxy properties (SFR, stellar mass, and specific SFR) and the ratio of nebular to the stellar A$_V$ (e.g., \citealt{Wild:2011aa,Price:2014aa,Reddy:2015aa}).

However, the two-component schematic model is built purely based on the OIR data, without the observational knowledge of the spatial distribution of dust. Through the analyses in \autoref{sec:sfr}, our data {on the six SMGs} suggest that the model becomes {incomplete} once the spatially resolved FIR data are considered, namely the majority of the dust attenuation cannot be traced via H$\alpha$ or OIR continuum only. A scenario that involves a third component, a centrally concentrated, extremely dusty component, appears more appropriate (\autoref{fig8}). Most of this third component of high dust concentration has a low filling factor and is likely not reflected by the rest-frame OIR and H$\alpha$. 

{Although this new scenario is based on the data of the six SMGs, it is also supported by other studies in the literature on OIR-selected galaxy samples; For example, even having spatially resolved estimates of dust attenuation}, \citet{Nelson:2019aa} found insufficient correction of their H$\alpha$ SFR in the central regions of a dusty galaxy at $z=1.25$, resulting also in a mismatch between the H$\alpha$ SFR density profile and that of the IR SFRs. Adding this component mitigates the tension between the A$_V$ derived from OIR photometry only and that from the UV-to-radio energy balance modeling. It also naturally explains the mismatch between the H$\alpha$-based SFRs and IR-based SFRs on our sample SMGs. {While we are lacking the measurements of nebular attenuation through Balmer decrement to completely verify this scenario on our sample SMGs, recent study by \citet{Scholtz:2020aa} on a sample of eight ALMA-detected, X-Ray selected AGN confirm that in six of their AGN that have constraints of Balmer decrement, the nebular A$_V$ values are indeed all larger than the stellar A$_V$ values. Together with the fact that they also find a similar FIR continuum to H$\alpha$ size ratio, the results of \citet{Scholtz:2020aa} support the three-component model.}

{While SMGs can be located on or above the massive-end ($\sim10^{11}$\,M$_\odot$) of the main sequence (e.g., \citealt{Michaowski:2012fr,da-Cunha:2015aa,Michaowski:2017aa}), due to their low space density the contribution of SMGs to the total cosmic star formation rate density is about 10-30\% \citep{Barger:2012lr,Swinbank:2014aa,Cowie:2017aa}. Therefore {if the three-component model were to be generalized to the other galaxies at $z\sim2$}, perhaps a more interesting question to ask is do typical $z\sim2$ star-forming galaxies that dominate the cosmic SFR density, and are on average less luminous in infrared have a similar FIR-to-H$\alpha$ spatial disparity? A few studies in the literature may give us some hints.

Through FIR measurements of both individual galaxies and stacked images of their ALMA data, it has been found that on average galaxies at $z\sim2$ are smaller if they have lower infrared luminosities ($\sim10^{12}$\,L$_\odot$ so SFR$\sim$100\,M$_\odot$ yr$^{-1}$; \citealt{Tadaki:2017aa,Fujimoto:2017ab,Fujimoto:2018aa}). In addition, they also found that the FIR sizes are a factor of 2-3 smaller than the rest-frame optical sizes for galaxies with lower infrared luminosities. In the meantime, using AO-aided SINFONI observations, \citet{Forster-Schreiber:2018aa} found that in their sample of $z\sim2$ star-forming galaxies that have a wide range of stellar masses ($2\times10^{9}-3\times10^{11}$M$_\odot$, median $\sim2\times10^{10}$\,M$_\odot$) and SFRs (10-650\,M$_\odot$ yr$^{-1}$, median $\sim80$\,M$_\odot$ yr$^{-1}$), the stellar sizes are consistent with the H$\alpha$ sizes to within about 5\% on average. These points of evidence suggest that a similar spatial disparity between FIR continuum and H$\alpha$ and OIR continuum may also exist in star-forming galaxies with infrared luminosity less than the SMGs. If true, the three-component dust model may be applicable to the massive star-forming galaxies at $z\sim2$ in general. }

\section{Summary}\label{sec:sum}
Using data from ALMA and near-infrared IFUs, we study the two-dimensional distributions of FIR continuum and H$\alpha$ for a sample of six $z\sim2$ SMGs. The objectives are to study their relative distributions and sizes, and investigate the impact of the results on issues such as the estimated star-formation rates, dust correction, dust distribution, and the SED modeling. We summarize our findings in the following:

\begin{enumerate}
	\item The sizes of H$\alpha$ are significantly ($>3$\,$\sigma$) larger than those of the FIR continuum in half of our sample SMGs (\autoref{fig3}). Across the sample the H$\alpha$ sizes are a median factor of ${2.0\pm0.4}$ larger than the FIR sizes.
	
	\item We find that the observed H$\alpha$-based SFRs are systematically lower than the IR-based SFRs by a median factor of $20\pm15$, and the factor is $4\pm2$ with attenuation correction applying the stellar A$_V$ to H$\alpha$. By adopting the most extreme attenuation correction provided by \citet{Calzetti:2000aa}, the difference is still a factor of $3\pm1$ (\autoref{fig4}).
	
	\item By plotting the one-dimensional SFR density profiles (\autoref{fig5}) we find that less than 50\% of the H$\alpha$ emissions come from outside the radii spanned by the dusty star-forming regions as traced by the FIR continuum. This is therefore not sufficient in explaining the differences found between H$\alpha$-based SFRs and the IR-based SFRs. Given the expected high attenuation in the central dusty regions, we postulate that the majority of the obscured star formation is not reflected through the extinction of OIR emissions including H$\alpha$. {It could be because that, in the scenario of foreground screen of dust, the FIR continuum and OIR emissions are spatially decoupled due to extremely high A$_V$ ($>$100). It could also be, in the opposite scenario of dust mixing with stars, because of the OIR emissions becoming optically thick due to high column density. A mixture of both scenarios is also possible.} 
	
	\item To understand the impact of spatial mismatches on the energy balance SED modeling, we compare the A$_V$ values derived from {\sc hyperz} using the OIR photometry alone and those from {\sc magphys}. We find that the A$_V$ values from {\sc magphys} are significantly higher in two SMGs (out of five; one source is excluded due to foreground contaminations), where the two have the highest IR luminosites among the sample. This suggests that the energy balance approach deduces a higher A$_V$ to account for the extra IR luminosities. For the three SMGs in which the A$_V$ values agree, we postulate that the contribution of the IR emissions from the birth clouds could be unrealistically high.
	
	\item Finally, considering the observed morphologies of H$\alpha$, OIR and FIR continuum, together with our findings about SFRs and A$_V$, we postulate that the dust distributions in SMGs, and {possibly} also in less IR-luminous $z\sim2$ star-forming galaxies, can be decomposed into three components; the diffuse ISM component obscuring the older stellar populations, the more obscured young star-forming H{\sc ii} regions, and the heavily obscured central regions with low filling factor in which the bulk of attenuation cannot be reflected through H$\alpha$ or OIR continuum.
	
\end{enumerate}

\begin{acknowledgements}
      {We thank the anonymous referee for the constructive feedback that improved the manuscript.} We also thank Richard Anderson for advices on the GAIA DR2 catalog. C.-C.C. is supported by an ESO fellowship program and is especially grateful to Tzu-Ying Lee (李姿瑩), who ensured the arrival of I-Fei Max Lee (李亦飛) during the time of this work. IRS and AMS acknowledge support from STFC (ST/P000541/1). JLW acknowledges support from an STFC Ernest Rutherford Fellowship (ST/P004784/2). MJM acknowledges the support of the National Science Centre, Poland, through the SONATA BIS grant 2018/30/E/ST9/00208. This paper makes use of the VLT data obtained through the following programs: 091.B-0920(E), 091.B-0920(C), 094.B-0798(B) and 096.A-0025(A). This paper also makes use of the following ALMA data: ADS/JAO.ALMA\#2012.1.00307.S, ADS/JAO.ALMA\#2015.1.01528.S, ADS/JAO.ALMA\#2016.1.00434, and ADS/JAO.ALMA\#2016.1.00735.S. ALMA is a partnership of ESO (representing its member states), NSF (USA) and NINS (Japan), together with NRC (Canada), MOST and ASIAA (Taiwan), and KASI (Republic of Korea), in cooperation with the Republic of Chile. The Joint ALMA Observatory is operated by ESO, AUI/NRAO and NAOJ. This research made use of Astropy,\footnote{http://www.astropy.org} a community-developed core Python package for Astronomy \citep{Astropy-Collaboration:2013aa,Astropy-Collaboration:2018aa}.  
\end{acknowledgements}

\end{CJK}

%
%

\bibliographystyle{aa}
\bibliography{bib} 

\end{document}